\documentclass[journal,twoside,web]{ieeecolor}

\usepackage[T1]{fontenc}         
\usepackage[utf8]{inputenc}      
\usepackage{jsen}
\usepackage[english]{babel}
\usepackage{csquotes}
\usepackage{graphicx}
\usepackage{float}
\usepackage{booktabs}
\usepackage{multirow}
\usepackage{wrapfig}
\usepackage{comment}
\usepackage[figuresright]{rotating} 
\usepackage{booktabs}  
\usepackage[table]{xcolor}
\usepackage{array}
\usepackage{xcolor}
\usepackage{amsmath}
\usepackage{tikz}
\usepackage{textcomp}
\usepackage{lipsum}
\usepackage[T1]{fontenc}
\usepackage{booktabs}
\usepackage{multirow}
\usepackage{siunitx}

\usepackage[backend=biber,giveninits=true,isbn=false,doi=false,
eprint=false,url=true,bibstyle=ieee]{biblatex}
\def\BibTeX{{\rm B\kern-.05em{\sc i\kern-.025em b}\kern-.08em
    T\kern-.1667em\lower.7ex\hbox{E}\kern-.125emX}}
\definecolor{abstractbg}{rgb}{0.89804,0.94510,0.83137}
\newcommand\acceptedtext{%
	\footnotesize This article has been accepted for publication in a future issue of this journal, but has not been fully edited. Content may change prior to final publication. \\
	Citation information: DOI 10.1109/JSEN.2022.3223297, IEEE Sensors Journal.}
\newcommand\acceptednotice{%
	\begin{tikzpicture}[remember picture,overlay]
		\node[anchor=north,yshift=-6pt] at (current page.north) {%
			\begin{minipage}{\textwidth}
				\center \acceptedtext
		\end{minipage}};
	\end{tikzpicture}%
}

\begin{document}

\title{
Sensing Technologies for Crowd Management, Adaptation,
and Information Dissemination in Public
Transportation Systems: A Review
}

\author{Donatella~Darsena,~\IEEEmembership{Senior Member,~IEEE,}
        Giacinto~Gelli,~\IEEEmembership{Senior Member,~IEEE,}
        Ivan Iudice, \\ and
        Francesco~Verde,~\IEEEmembership{Senior Member,~IEEE}
\thanks{
Manuscript received December 1, 2021;
revised April 30, 2022, September 17, 2022, and October 21, 2022;
accepted November 14, 2022.
The editor coordinating the review of
this manuscript and approving it for publication was Dr.~Prosanta Gope.}
\thanks{
D.~Darsena is with the Department of Engineering,
Parthenope University, Naples I-80143, Italy (e-mail: darsena@uniparthenope.it).
I.~Iudice is with Italian Aerospace Research Centre (CIRA),  Capua I-81043, Italy
(e-mail: i.iudice@cira.it).
G.~Gelli and F.~Verde are both with the Department of Electrical Engineering and Information Technology, University Federico II, Naples I-80125,
Italy [e-mail: (gelli,f.verde)@unina.it].}
\thanks{D.~Darsena, G.~Gelli, and F.~Verde are also with
National Inter-University Consortium for Telecommunications (CNIT).}
}

\IEEEtitleabstractindextext{%
\fcolorbox{abstractbg}{abstractbg}{%
\begin{minipage}{\textwidth}%
\begin{wrapfigure}[18]{r}{2.2in}%
\includegraphics[width=2.2in]{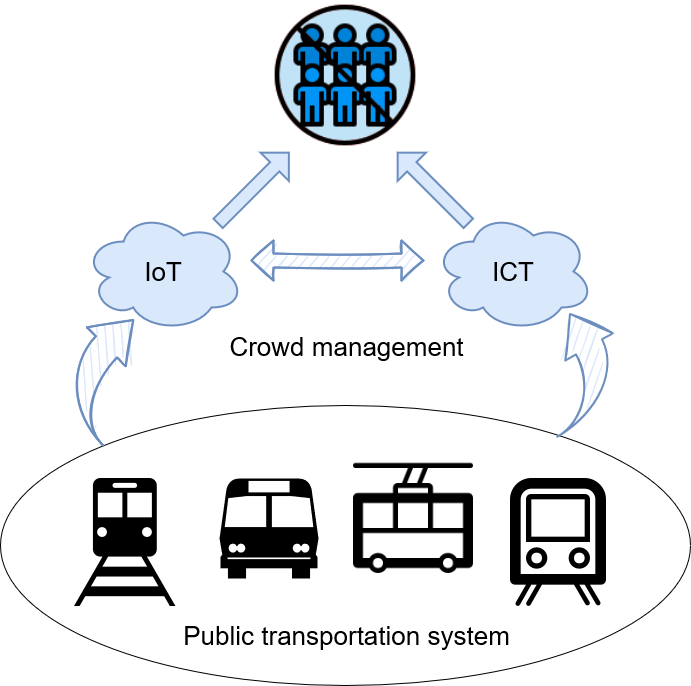}%
\end{wrapfigure}%

\begin{abstract}
Management of crowd information
in public transportation (PT)
systems is crucial, both to foster sustainable mobility, by
increasing the user's comfort and satisfaction
during normal operation,
as well as to cope with emergency situ\-ations, such as
pandemic crises, as recently experienced
with COVID-19 limitations.
This paper presents a taxonomy and review of
sensing technologies based on Internet of Things (IoT)
for real-time crowd analysis, which can be
adopted in the different segments
of the PT system
(buses/trams/trains, railway/metro stations,
and bus/tram stops).
To discuss such technologies
in a clear systematic perspective, we
introduce a reference architecture for crowd management,
which employs modern information
and communication technologies (ICT) in order to:
(i) monitor and predict crowding events;
(ii) implement crowd-aware policies
for real-time and adaptive operation
control in intelligent
transportation systems (ITSs);
(iii) inform in real-time the users
of the crowding
status of the PT system,
by means of electronic displays installed inside vehicles
or at bus/tram stops/stations,
and/or by mobile transport applications.
It is envisioned that the innovative crowd
management functionalities
enabled by ICT/IoT sensing technologies
can be incrementally implemented
as an add-on to state-of-the-art ITS platforms, which are
already in use by major PT companies operating
in urban areas.
Moreover, it is argued that,
in this new framework,
additional services can be delivered
to the passengers,
such as, e.g., on-line ticketing,
vehicle access control and reservation
in severely crowded situations,
and evolved crowd-aware route planning.
\end{abstract}

\begin{IEEEkeywords}
Adaptive systems,
COVID-19,
communication systems,
crowd management,
intelligent transportation system (ITS),
Internet of Things (IoT),
public transportation (PT) systems,
sensing technologies,
sustainable mobility,
smart cities.
\end{IEEEkeywords}
\end{minipage}}}

\pubid{\begin{minipage}[t]{\textwidth}\ \\[10pt]
		\centering\footnotesize{1558-1748 \copyright 2022 IEEE.
			Personal use is permitted, but republication/redistribution requires IEEE permission.\\
			See https://www.ieee.org/publications/rights/index.html for more information.}
\end{minipage}} 
\maketitle
\acceptednotice

\section{Introduction}
\label{sec:intro}

\IEEEPARstart{S}{ince} 2002, the
European Commission has promoted
across Europe and beyond
a campaign supporting the use of public
transportation (PT) systems  as ``a safe, efficient,
affordable, and low-emission mobility
solution for everyone'' \cite{EUweek}.
To cope particularly with PT limitations in urban areas,
a key role is expected to be played by
intelligent transportation systems (ITSs)
\cite{Dimitrakopoulos2020},
which leverage
information and
communication technologies (ICT)
to enable automated collection of transportation data,
used to make transport safer, more
efficient, more reliable,
and more sustainable.

Two different classes of
spatio-temporal data are available in state-of-the-art ITSs for
PT:
(i) \emph{vehicle data},
such as location and speed,
which are obtained (usually in real-time)
by means of automated vehicle location (AVL)
\cite{Hounsell2012} systems,
mainly based on satellite
localization techniques; and
(ii) \emph{passenger data},
such as the number of passengers boarding a bus/train
or entering a station,
which can be obtained  by means of automatic
passenger counting (APC) \cite{Pinna2010}
or automatic fare collection (AFC) systems \cite{Opurum2012}.
Such data are commonly used as inputs
for optimization and planning strategies
for PT systems, which
are surveyed in \cite{Iliopoulou2019}
at four different levels:
strategic, tactical, operational, and real-time.
In particular,
it is observed in \cite{Iliopoulou2019}
that lack of passenger
arrival information, especially in real-time,
is a limiting factor for accurate
studies.
In many ITSs, indeed,
real-time location data are available only for
vehicles, which are used both to provide
trip information to passengers
and for medium-to-long term management
and monitoring of the service.

With the advent of Internet of Things
(IoT) technologies, fine-grained
and real-time passenger data collection
is becoming a feasible task, especially in
smart cities \cite{Habibzadeh2017}.
Indeed, the pervasive use of mobile and portable
devices, equipped with different
sensors, allows one to gather huge quantities
of data in urban scenarios \cite{Cardone2014,Gallo2014,Kaiser2018,Khalid2019},
which can be used for different
applications and tasks \cite{Sheng2013}.
In particular, the fifth generation (5G)
of cellular networks
is potentially able to
support massive IoT connections
\cite{Palattella2016},
where billions of smart devices
are connected to the Internet and can be
easily located and tracked; these features will be further
extended by forthcoming sixth generation (6G) networks \cite{Guo2021}.

Recently, some researchers
have argued that
the quality of service (QoS)
perceived by PT users,
as well as their travel satisfaction/quality of experience (QoE),
are seriously
affected by \emph{crowding}
\cite{Cantwell2009,Tirachini2013,Drabicki2021}.
To cope with this issue,
it is required to acquire in {real-time}
reliable and capillary information
about the crowding status of
PT rail or road vehicles
(e.g., buses, trams, and trains)
and the related access
infrastructures (e.g., bus/tram stops and
metro/railway stations).
Indeed, it is stated in \cite{SanchezMartinez2016} that
``the availability of real-time passenger
demand data can significantly
improve the performance of control models
in case of overcrowding''.

Motivated by the previous needs, new modeling,
planning and management
strategies that collect real-time crowding
data and use them to improve QoS/QoE
in PT systems
are appearing in the
literature \cite{Drabicki2021,Handte2014,Farkas2015,Handte2016,Wang2021c},
and will be referred in the following
as \emph{crowd management} \cite{Franke2015,Boukerche2019}.
Crowd management systems are
composed  \cite{Boukerche2019}
by \emph{crowd analysis/monitoring} and
\emph{crowd control} components:
the former includes a network of physical sensors aimed
at detecting crowds and estimating
their parameters,
whereas the latter includes prediction,
decision making and control strategies
aimed at managing the crowd events.
Crowd management
should not be confused with \textit{crowdsourcing/crowdsensing} \cite{Nash2018}.
In PT systems, crowdsourcing (also referred to as
``sensing by the crowd" approach)
is based on reporting activities by which
passengers provide their suggestions and feedback, as well as announce
problems to a PT company or, even, provide a
resource or create a product, e.g., peer-to-peer services.
Relying on crowdsourced data, PT companies work together with
passengers to solve a problem or jointly plan public transit by finding and
developing solutions aligned with PT user preferences.
Crowd management is instead based on the different principle of
``sensing the crowd", according to which
environmental data collected through networks of IoT sensor devices and/or user terminals
are shared with PT companies, which
analyze such data for forecasting, choosing
among several possible alternative options, as well as
ensuring robust and safe operation of PT systems.

The problem of crowd management in PT systems
has emerged dramatically during coronavirus disease (COVID-19)
pandemic, which have first spurred our interest to this topic
(see Section~\ref{sec:mobility} for a discussion).
During the acute outbreak phase,
overcrowding of buses and trains
needed to be strictly avoided
to protect people from contagion, which required
emergency measures, such as, e.g.,
limiting to $50\%$ the service capacity.
These measures are generally not sustainable
in the long-term, since they shift a significant
portion of passengers to private transportation.
Moreover, the problem of overcrowding in PT
systems must be tackled in a smarter and
more structural way,
since many experts predict that,
in the future, recurrent waves of
pandemic outbreaks could become the
norm rather than the exception.

Although state-of-the-art AVL/APC/AFC systems collect
a large amount of data
about vehicles and passengers
in ITS systems,
often they are not suitable to perform real-time
crowd monitoring and control.
However, many ICT/IoT sensing technologies
for crowd monitoring are already available
or will be available soon
in smart cities \cite{Habibzadeh2017}.
A recent special issue of this journal
\cite{Jolfaei2021} reports
the cutting-edge advances
and ICT technologies pertaining to
the seamless integration of
sensors with the
transportation infrastructure.
The main focus of the contributions
in \cite{Jolfaei2021} is on
sensing technologies for \emph{private}
transportation systems, oriented to autonomous driving
\cite{Lian2021,Yang2021},
intelligent fault detection
\cite{Singh2021,Yan2021},
electric charging optimization
\cite{Ejaz2021},
road condition monitoring \cite{Kortmann2021,Mishra2021},
precise fleet management \cite{Chen2021},
speed detection and accident
avoidance \cite{Peng2021,Wang2021a}:
in the same issue, less attention is
dedicated to the integration of
sensing technologies within PT systems.
Moreover, sensing techniques for crowd monitoring are
not specifically discussed.


\begin{table*}[!t]
\defaultaddspace 4pt
\centering
\caption{Comparison among surveys related to crowd management in PT systems}
\begin{tabular}{p{12mm}p{14mm}p{24mm}p{45mm}p{60mm}}
\toprule
{\bf Ref.}   & {\bf Reference}     & {\bf Considered} & {\bf Scenarios} & {\bf Sensing technologies } \\
             & {\bf architecture}  & {\bf aspects}    &                 &    \\
\midrule
\cite{Pinna2010}
& No
&  Crowd analysis
&  Passenger counting inside PT vehicles
&  Infrared sensors, treadle mat sensors, and weigh-in-motion systems
\\ \addlinespace
\cite{Kaiser2018}
&  Yes
&  Crowd analysis
&  PT systems are mentioned but
no scenario is discussed in detail
&  Call detail records, origin-destination matrices, GPS,
Bluetooth, WiFi, RFID, video, social networks,
activity trackers, satellite, UAV,
and census data
\\ \addlinespace
\cite{Tirachini2013}
&  No
& Crowd control
&  Inside buses and trains, inside stations
& Physical sensing is not discussed
\\ \addlinespace
\cite{Handte2016}
&  No
&  Crowd management
&  Inside buses
&  WiFi, GPS, and cellular system (3G)
\\ \addlinespace
\cite{Boukerche2019}
& Yes
& Crowd management
& Related to PT systems but
no scenario is discussed in detail
& CCTV systems, presence sensors, RFID,
tracking mobile devices, and smartphones
\\ \addlinespace
\cite{Grgurevic2022}
&  No
& Crowd analysis
& Passenger counting Inside PT vehicles
& Optical sensors, pressure sensors, computer vision, and WiFi
\\ \addlinespace
\cite{Cecaj2021}
&  No
&  Crowd analysis and prediction
&  PT systems are mentioned but
no scenario is discussed in detail
& WiFi, cellular systems (4G/5G), social media, cameras, and taxi trajectories
\\ \addlinespace
\cite{Xie2020}
&  Yes
&  Crowd prediction
&  Prediction of the urban flow (e.g. the traffic of crowds, vehicles, and bikes)
& Physical sensing is not discussed
\\ \addlinespace
\cite{Yuan2022}
&  Yes
&  Crowd management
&  PT systems are mentioned but
no scenario is discussed in detail
& Physical sensing is not discussed
\\ \addlinespace
\cite{Tyagi2020}
&  Yes
& Crowd analysis
&  PT systems are mentioned but
no scenario is discussed in detail
& Physical sensing is not discussed
\\ \addlinespace
\cite{Gao2020}
&  Yes
& Crowd analysis
&  PT systems are mentioned but
no scenario is discussed in detail
& Physical sensing is not discussed
\\ \addlinespace
\cite{Gkiotsalitis2021a}
&  No
&  Crowd control
&  Bus and tram stops
& Physical sensing is not discussed
\\ \addlinespace
This paper
& Yes
& Crowd management
& Train and bus/tram vehicles, railway/metro stations, and bus stops
& Infrared sensors, pressure/load sensors, optical and thermal cameras,
LiDAR, acoustic/ultrasound sensors, RFID and NFC devices, Bluetooth, WiFi, and cellular systems
(4G/5G)
\\ \bottomrule \\
\end{tabular}
\label{tab:tab-comparison}
\end{table*}


\subsection{Contribution}

In this paper, we argue that, although
many sensing technologies for crowd monitoring
are already available,
the diffusion of crowd management techniques in modern PT systems
is hindered by the lack of a structured framework and reference architecture.
Motivated by this
observation, we pursue three
main goals in this paper:
(i) to present a survey and taxonomy
of crowd monitoring technologies for PT systems
based on ICT/IoT technologies;
(ii) to discuss their adoption into a reference architecture,
aimed at integrating
state-of-the-art ITSs
(already available in many PT systems in urban areas)
with the new crowd management functionalities;
(iii)
to highlight a series of challenges opened up by
the proposed reference architecture
that need to be investigated, by also
indicating what could be the best way to address them.
Although some reviews regarding techniques for
crowd monitoring inside PT vehicles
exist (see \cite{Pinna2010,Grgurevic2022}),
to the best of our knowledge, this is the first review
addressing sensing technologies for crowd management
in \emph{all} the different segments
of the PT systems.
In this respect, Table~\ref{tab:tab-comparison} enlists all the related surveys in
crowd management in PT systems, by comparing them
with the review paper at hand.

In our vision,
the crowd management functionalities
are built upon a distributed IoT subsystem,
composed by a capillary network of heterogenous
active/passive sensors, aimed at monitoring
passenger crowding inside
buses, trams, and trains, at bus/tram stops,
and in railway/metro stations.
By means of a communication
infrastructure, the acquired measures are transmitted
in real-time to a smart subsystem,
which performs crowd control functionalities.
Crowding information can also be reported
(in aggregated or anonymized form, for privacy concerns)
to PT users, by means of displays installed inside
vehicles or at stations/bus stops, or
through mobile transport apps (e.g., Moovit or
proprietary operator
applications).

In the proposed reference architecture,
real-time knowledge of crowding
data can be used
by PT operators
for fast or even \emph{proactive}
adaptation of some service features
(e.g., vehicle holding,
stop-skipping, overtaking, limited boarding,
speed changing, short turning), in order to cope with
spatially and/or temporally localized crowding situations,
which cannot be tackled by conventional (statistical)
tools used in transportation system design,
such as, e.g., origin-destination (O-D) flow analysis.
The new crowd management functionalities
essentially achieve two scopes:
(i) to improve the QoS/QoE of passengers, thereby
fostering PT system usage;
(ii) to allow for safe PT usage
during exceptional events like a
pandemic outbreak, such as COVID-19.

\subsection{Paper organization}

Section~\ref{sec:mobility}
highlights the adverse impact of COVID-19
pandemic outbreak on sustainable mobility and
PT systems.
In Section~\ref{sec:crowd-management},
the main aspects of
crowd management in PT systems are discussed.
In Section~\ref{sec:architecture},
a crowd management reference
architecture is proposed.
A taxonomy and review of the sensing solutions for
crowd management and their application in different PT scenarios
is presented in Section~\ref{sec:SASS}.
Innovations and advantages
provided by the
adoption of the new crowd management
functionalities are highlighted
in Section~\ref{sec:i-a}.
The main challenges and gaps
related to the introduction of
crowd monitoring management in ITS systems
are discussed in Section~\ref{sec:chall}.
Finally, conclusions are drawn
in Section~\ref{sec:concl}.


\section{Sustainable mobility in the COVID-19 era}
\label{sec:mobility}

During the last years,
the transport
sector and mobility -- in particular PT systems in urban
areas -- have been seriously affected by
COVID-19 pandemic.
A survey \cite{Ipsos2020} carried out in China
in 2020 estimated that, as a consequence
of the outbreak, the use of private cars
will be roughly doubled,
increasing from $34\%$ to $66\%$,
whereas the use of public transports
(buses/metros) will be more than halved,
dropping down from $56\%$ to $24\%$.
Furthermore, due to the lack
of trust in PT systems,
more than $70\%$ of the surveyed
people not owning a car declared their intention
to buy a new one, with negative
consequences on the environment (landscape and air pollution)
in urban areas.
Other recent studies (see, e.g., \cite{Gutierrez2020,Aloi2020})
have highlighted that COVID-19 pandemic
has seriously discouraged the use of PT systems.

To counteract the shift to
private car usage during COVID-19
pandemic, national governments have implemented
different strategies.
A widespread measure \cite{Campisi2020} has been
to favor the use of individual
sustainable mobility and micromobility means,
such as bikes, electrical scooters, and segways,
by deploying the related
infrastructures (bike lanes)
or empowering vehicle sharing services,
which can shift to this transport
mode a certain percentage of
short and medium-distance trips.

However, owing to the large number of
passengers carried by PT systems in urban areas,
it is of utmost importance  to adopt
measures aimed at
safe and reliable PT usage in such a scenario.
During the acute phase of the outbreak,
severe anti-COVID-19 measures were adopted \cite{Tirachini2020}
to minimize the contagion risk, such as
back-door  boarding,  cashless  operations,
frequent sanitization of vehicles
and stations, enforcing social distances,
limiting the service capacity,
and requiring the passengers to wear
face masks.
Other anti-COVID-19 measures were
applied \cite{Tirachini2020,Gkiotsalitis2021} to PT system operations, such as
modifying timetables, frequencies, paths,
leveraging modal integration, and so on.
Unlike other countries,
the Land Transport Authority (LTA) of Singapore
has made adjustments to train
frequencies to reduce crowding
on commuters' lines, increasing
the peak-period frequency for trains from once
every five minutes to once every three minutes \cite{LTA2}.
Some of these measures, like
increasing PT service frequency or introducing extraordinary trips
to compensate for the reduced vehicle capacity,
are seen
by PT companies as effective,
but not sustainable in the long term,
due to the limited number
of drivers and vehicles and the increased
operational costs \cite{Coppola2020}.

Generally speaking, the COVID-19 pandemic
has pushed towards a critical rethinking of many
economical, social, and cultural habits, not only those
related to sustainable mobility.
A plethora of innovative solutions have been proposed to cope with
this new challenge, many of them employing
ICT/IoT technologies.
In \cite{Chamola2020}, the use of new technologies,
such as IoT, unmanned aerial vehicles
(UAVs), artificial intelligence (AI),
blockchain, and 5G, has been considered for managing the impact of
COVID-19 in health applications.
In \cite{Nguyen2020a,Nguyen2020b}
a review of technologies
for social distancing has been provided, with emphasis
on wireless technologies for positioning, including crowd detection
and people density estimation.

As far as sustainable mobility is concerned,
a review of the PT planning literature
can be found in \cite{Gkiotsalitis2021}
from the perspective of the changes in demand patterns
and limited capacity requirements associated with the
COVID-19 pandemic crisis.
It is evidenced that, besides
reducing the service capacity to adhere to
physical distancing measures,  PT service providers
worldwide have resorted to limiting service span
in order to reduce operational costs as a consequence
of the reduction in catchment area, by canceling
certain services or closing some stations.
Moreover, it is pointed
out the importance to develop and deploy methods that
are able to maintain the functionality of PT systems,
while minimizing the public health risks;
it is suggested that
some changes in service provision can be made
at the {tactical planning phase}, by modifying
timetables and/or service frequencies.

Some recent studies have explored
the possibility to rely on ICT
to organize and
facilitate human mobility during the pandemic.
In \cite{Asad2020}, the Authors propose to use
a machine-learning (ML) based approach to trace
daily train travelers in different age cohorts of $16$--$59$ years
(i.e., the less vulnerable age-group) and over
$60 $ (i.e., the more vulnerable age-group) in order to recommend
certain times and routes for safe traveling.
In this work, many ICT technologies, such as WiFi,
Radio-Frequency IDentification (RFID),
Bluetooth, and Ultra WideBand (UWB), are employed.
Using a dataset of the London underground and overground network,
different ML algorithms are compared in \cite{Asad2020}
to properly classify different age group travelers,
showing that the Support Vector Machine (SVM) approach
performs better to predict the mobility
of travelers and achieves high accuracy (more than $80\%$).

In \cite{Reigber2006},
a comprehensive review
on human mobility research
using big data is carried out:
big data collected thanks to the pervasive use of
ICT/IoT can help, indeed, to discover the relationships between
human mobility and resource use, thus entailing great
opportunities for smart city development.
In \cite{Rahman2021}, instead, the Authors pursue
the objective of identifying data sources and ML approaches
to properly estimate the impact of COVID-19 on human
mobility reduction.
In particular,
the consequences of the pandemic on mobility
patterns of urban populations are investigated
in \cite{Rahman2021}, by quantifying
even the impact of mobility reduction
on improving air quality in urban areas.


\begin{figure*}[!t]
\centerline{\includegraphics[width=0.90\columnwidth]{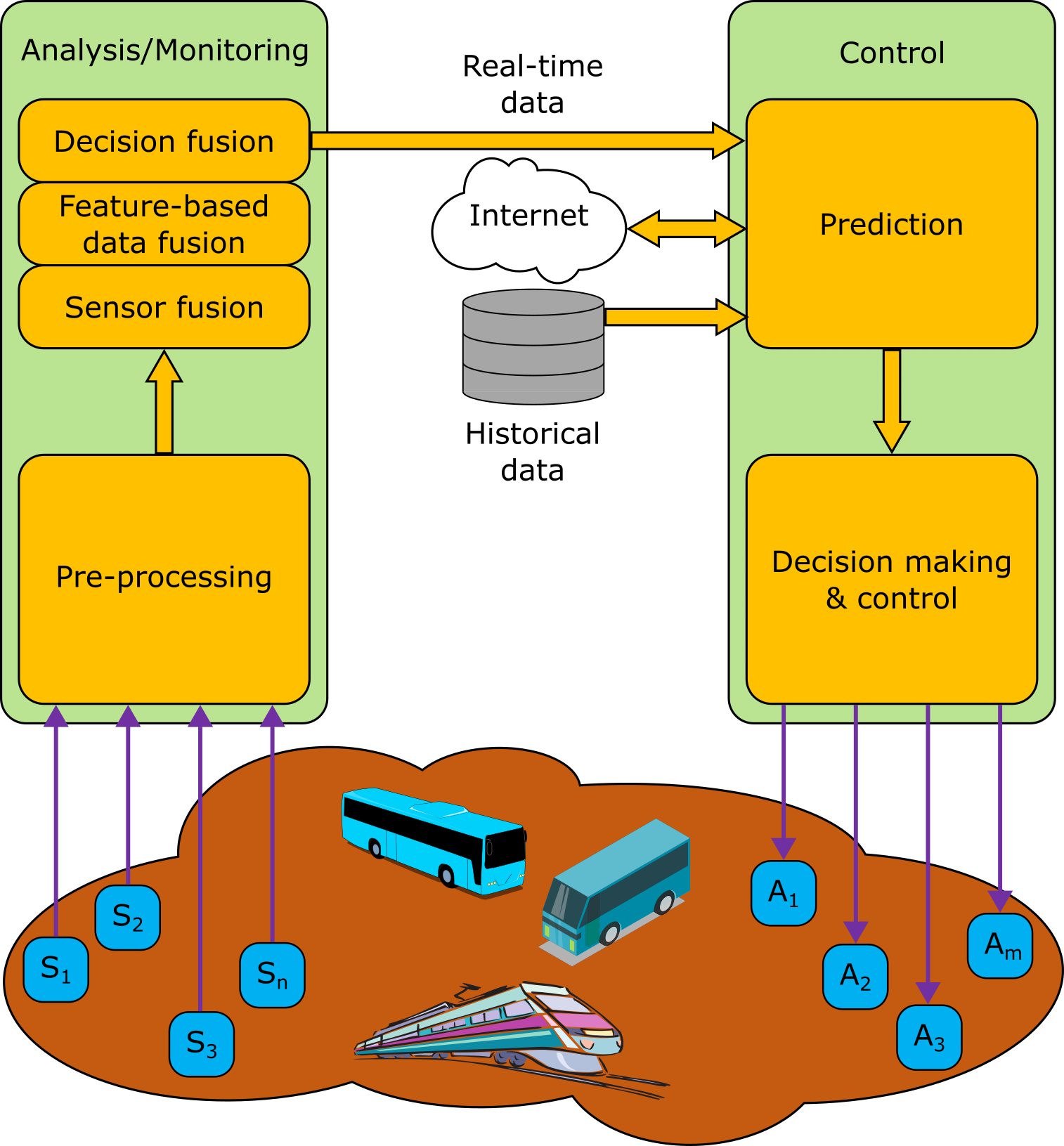}}
\caption{Main architecture of the crowd analysis/monitoring and crowd control mechanisms.}
\label{fig:architecture2}
\end{figure*}

\section{Crowd management in PT systems}
\label{sec:crowd-management}
\label{sec:concept}

The main architecture of a  crowd management system
for PT applications is schematically depicted in Fig.~\ref{fig:architecture2}.
A certain number of physical sensors $\text{S}_1, \text{S}_2, \ldots, \text{S}_n$
are used to collect all the relevant information regarding crowd events.
After a suitable pre-processing of the collected information
(e.g., dimensionality reduction, normalization, interpolation, denoising, and
so forth), the output data are processed by the data-fusion system.
Data-fusion can conceptually be divided into three layers \cite{Li2021}:
sensor fusion, feature-based data fusion, and decision fusion.
Sensor fusion includes data classification, object refinement (e.g.,
spatio-temporal information alignment, correlation, clustering,
grouping, state estimation, error elimination, and reduction),
positioning, and recognition. The output of the sensor-function
layer has a consistent data structure.
Data feature fusion helps to reduce the requirements of application services
for system storage resources and computing performance, and it can provide additional
complete and in-depth features.
Decision-level fusion is aimed at detecting crowds and estimating
their parameters in real-time.
Prediction of crowd events and their manage is obtained
by elaborating not only the output data of the
decision-fusion layer, but also other information arising from
the Internet and historical databases. Control actions are
finally implemented through a certain number of actuators
$\text{A}_1, \text{A}_2, \ldots, \text{A}_m$.

A general introduction to
crowd management in transportation systems
is provided in \cite{Boukerche2019},
where some limitations of
state-of-the-art ITSs are highlighted,
and the potentials of
the new approach are discussed, together with a
brief introduction to
crowd analysis/monitoring techniques.
In \cite{Tirachini2013},
different aspects of passenger crowding in PT systems
are discussed, related to demand,
supply, and operations, including effects
on route and bus choice,
as well as passengers' wellbeing.
Table~\ref{tab:tab-crowd-management}
subsumes some relevant works on crowd management discussed
in the following,
focusing in particular to those reporting
numerical performance achievements in realistic scenarios.

The benefits of \emph{real-time crowding
information} (RTCI) dissemination on
passenger travel choices have been discussed
and assessed by
simulations in \cite{Drabicki2021,Drabicki2017,Drabicki2022}.
In particular, it is shown in \cite{Drabicki2017,Drabicki2022}
that providing RTCI at bus stops
might help reduce the deleterious
``bus bunching'' effect \cite{Bellei2010}.
In \cite{Drabicki2021}
a complete framework
for RTCI modeling
in PT systems is introduced,
which incorporates RTCI
in a dynamic path choice model: the new methodology is tested on
a simplified model of the urban PT system
of Krak\'{o}w, Poland, showing that RTCI dissemination
contributes to a more efficient distribution
of passenger loads in the PT network, improving travel
confort and reducing waiting time by about 30\%.

A field application of crowd management strategies
to PT systems is considered in
\cite{Handte2014,Handte2016}.
The proposed solution, tested in the municipal
bus infrastructure of Madrid, Spain,
estimates the current number of passengers
in a bus by exploiting the properties
of the existing WiFi connections
(see Section~\ref{sec:device-aided})
and incorporates such information
in a bus navigation system, which is
capable of giving crowd-aware
route recommendations.
Well before the COVID-19 outbreak,
the LTA in Singapore has been collecting AFC data \cite{LTA1}
to help identify commuter hotspots,
which enables them to better
manage bus fleets and commuter demand, achieving
remarkable service improvements, such as
92\% reduction in the number of bus services with crowding issues,
and 3- to 7-minute reduction in the average waiting time on popular bus services
\cite{LTA3}.
It is worth noting that recently
also Google started acquiring and disseminating crowding
information to the users
\cite{Hasaballah2019,Hasaballah2019a,Tholome2021}.


%
\begin{table*}[!t]
\centering
\caption{Representative performances of crowd management solutions for PT systems}
\begin{tabular}{p{14mm}p{15mm}p{35mm}p{30mm}p{63mm}}
\toprule
{\bf Ref.} & {\bf Scenarios} & {\bf Approach} &{\bf Validation} & {\bf Performance achievements} \\
\midrule
\cite{Drabicki2021,Drabicki2017,Drabicki2022}
&  PT network
&  RTCI dissemination
&  Simplified version of PT network of Krak\'{o}w, Poland  
&  More efficient distribution of passenger loads, reduction of the
bus bunching effect, improvement of travel confort
and reduction of waiting time by about 30\%
\\ \addlinespace
\cite{Wang2021c}
& Bus network
& RTCI dissemination and bus holding
& Calibration of the model by a survey in Dalian City, China
& Reduction of bus headway and single trip time by 20\%,
reduction of in-vehiche crowdedness by up to 25\%
\\ \addlinespace
\cite{Jenelius2020,Jenelius2020a}
& Metro network
& RTCI dissemination,
real-time crowd estimation and prediction
& Metro line in Stockholm, Sweden
& Prediction accuracy improved
from 70--80\% to 80--95\%
\\ \addlinespace
\cite{LTA1,LTA2,LTA3}
& PT network
& Management of bus fleet and commuter demand
& LTA network in Singapore  
& Reduction of 92\% in the number of bus services with crowding issues,
reduction of 3- to 7-minute in the average waiting time on popular bus services
\\ \addlinespace
\cite{Bai2021}
& Metro line
& Optimization of timetables to reduce
platform over-crowding and passenger waiting time
& Bejing metro data  
& Reduction of the PWT by 17.18\% (off-peak hours)
or 3.22\% (peak hours), reduction of the number of passengers
on platforms by 44.5\% (off-peak hours)
and 9.5\% (peak hours)
\\ \addlinespace
\cite{Arabghalizi2020}
& Bus network
& Predictive model for bus crowding based on ML
& Bus system of Pittsburgh-area, USA
& Prediction accuracy improved by 8x over baseline models
\\ \addlinespace
\cite{Munoz2018}
& Metro line
& Optimization of the passenger flows by installing
unidirectional gates
at the platforms
& Metro line 4 in Santiago, Chile
& Increase of the transport capacity by 5\%, reduction of the travel time by 6.5\%,
positive passenger feedback
\\ \addlinespace
\cite{Luan2022}
& PT railway network
& Joint optimization of train scheduling
and passenger routes based on in-vehicle crowding
& Urban railway network in Z\"{u}rich, Switzerland 
& Reduction of the passenger in-vehicle travel time by 39\%
\\ \bottomrule
\multicolumn{4}{p{100mm}}{PWT = Passenger Waiting Time, EWT = Extended Waiting Time}
\end{tabular}
\label{tab:tab-crowd-management}
\end{table*}


\subsection{Crowd prediction}

Estimating and predicting (in space and/or time)
some features of a human
``crowd'' in indoor and
outdoor locations is an active research topic,
with many applications, including
surveillance and security,
situation awareness, emergencies, and crowd
management \cite{Cecaj2021}.
In transport applications, the feature of interest
is the number of components of
the crowd and/or its density.

Crowd prediction algorithms for transport applications
can be classified
on the basis of the
prediction horizon into
\textit{short-term} (less than $60$ minutes)
and \textit{long-term} ones \cite{Bai2021}.
Moreover, they can be
classified, on the basis of the prediction methodology,
into \emph{model-based} methods
or \emph{data-driven} ones.

Model-based methods include time-series analysis,
regression modeling, hidden Markov models
\cite{Fang2019},
and Kalman filtering models \cite{Zhang2017}.
Due to highly nonlinear and random nature of crowds, data-driven approaches
have recently gained significant attention, including ML and deep
learning (DL) techniques. Several works consider ML-based approaches
for crowd management (see, for example, \cite{Xie2020,Luca2021}).
In \cite{Yuan2022} the authors provide a thorough survey on ML techniques
for intelligent transportation systems.

Although ML-based approaches can achieve good results in crowd flow prediction, several reasons pushed researchers to adopt DL-based methods, such as, above all, the ability to automatically extract relevant patterns from unstructured and heterogeneous data. Differently from ML ones, indeed, DL approaches do not require manually extracted or handcrafted features, but can automatically extract the relevant features from the raw data collected by the sensors, process them and make the subsequent decision.
In \cite{Tyagi2020} the reference structure of a crowd counting technique based on a
convolutional neural network (CNN) is reported, whereas in \cite{Gao2020} crowd counting techniques are classified, according to the network property, into basic, scale aware models, context aware models, and multitask models.
In some cases,  simulation
tools to predict traffic flow, like BusMezzo \cite{Toledo2010}
or SUMO \cite{Lopez2018},
have been considered as well.

Although crowd/traffic prediction and mobility forecasting
are considered in several papers,
mostly devoted
to road congestion management in urban transportation,
their application to PT systems is a relatively new topic:
some recent studies
are \cite{Zhang2017,Nuzzolo2016,Arabghalizi2020,Jenelius2020,Jenelius2020a}.
In \cite{Arabghalizi2020}, a predictive model for bus crowding is proposed
and tested on a AVL/APC
dataset taken from the Pittsburgh-area bus using
ML techniques.
In \cite{Jenelius2020} a data-driven
approach is considered to
perform car-specific, metro, and
train crowding prediction,
aimed at
providing accurate
in-vehicle or station RTCI.

Real-time crowd estimation in \cite{Jenelius2020} is based on
load sensors (see Section \ref{sec:load-sensors}), which are
used, together with historical data, to perform
crowd prediction, whose accuracy is tested
with data gathered on a metro line in
Stockholm, Sweden.
It is shown that real-time load data significantly
improve prediction accuracy (from between 70\% to 90\% to between
80\% to 95\%).
In \cite{Jenelius2020a}
a framework for personalized (i.e., user-specific)
crowd prediction is proposed, which takes into account
not only loading data, but also other parameters that
affect user confort, such as seat availability,
expected travel time standing, and excess
perceived travel time
(compared to uncrowded conditions).
In \cite{Noursalehi2021}
a system  providing real-time short-term
crowd predictions on trains and platforms is proposed,
which uses both real-time
AFC and O-D historical data.


\subsection{Crowd control}

Crowd information
can be incorporated in PT design, optimization,
and control at different levels.
Besides some approaches showing
the benefits of  RTCI dissemination to the passengers
\cite{Drabicki2021,Drabicki2017,Drabicki2022,Jenelius2020},
operator-based crowd control in PT systems is still
at an early stage of development.
Crowding data can be used at the
strategic and tactical
planning level in several fields, such as
increased services, vehicle capacities,
networks expansion, headway and
timetable optimization
(see e.g. \cite{Sohn2013,Munoz2018}
for solutions applied
to the metro/subway scenario).
In \cite{Gkiotsalitis2016} a sequential heuristic
method is introduced for re-scheduling
the timetables of demand-responsive public transport
modes in near-real time.
The approach was tested on data of Stockholm
bus service, with reference to a figure of merit
called EWT (Excess Waiting Time).
Due to the schedule changes, the operational
performance of bus services
demonstrated a remarkable service-wide EWT
improvement (e.g., the service-wide EWT is reduced from 1.13 to 0.75 mins
for the line 1).

In \cite{Bai2021} a timetable optimization method
aimed at reducing the passenger waiting time (PWT)
in metro scenarios is proposed,
employing a Genetic Algorithm (GA)
integrated with the Interior-Point Algorithm (IPA).
The proposed method
is tested by simulation on data of
the Bejing Metro, showing
that the PWT under the optimized
timetable is reduced at best by
17.18~\% in off-peak hour and by 3.22~\%
in peak hour in comparison with
the standard timetable.
Moreover, the method reduces the peak number
of passengers on platforms by 44.5~\% during off-peak hours
and by 9.5~\% during peak hours.

In many studies,
the effects of crowding are represented
by an additional in-vehicle travel time multiplier,
so-called \textit{crowding penalty}
\cite{Tirachini2013}, which is typically
estimated from historical data.
A recent discussion of crowd-related
studies in PT railway systems
is provided in \cite{Luan2022}, which also
proposes a control model for joint optimization
of train scheduling and passenger routes
taking into account in-vehicle crowding
on the basis of O-D data.
Real-time crowd control
techniques are less explored,
mainly due to the lack of availability of
fine-grained crowding data.
Common measures to cope with service
irregularities in scheduled PT systems
are vehicle holding,  stop-skipping, overtaking,
limited boarding, speed changing and
short turning \cite{Cao2019,Gkiotsalitis2021a}.
All these strategies can further benefit from the availability
of real-time crowding information, both inside vehicles and
at the access infrastructure
(e.g., bus/tram stops and railway/metro stations).


\section{A reference architecture for crowd management}
\label{sec:architecture}

From the previous discussion,
it is apparent that integration of
ICT/IoT sensing technologies into PT systems for crowd management
is fragmentary and their
potentials are not fully exploited to date.
To bring together these
technologies in a systematic and common scenario,
we introduce a \emph{crowd management reference architecture}
for ITS, whose scope is to integrate/augment
the ITS system already
available in a urban PT system
with the new {crowd management} functionalities,
aimed at smart and proactive
control and reduction of passenger crowding.


\begin{figure*}[!t]
\centerline{\includegraphics[width=\columnwidth]{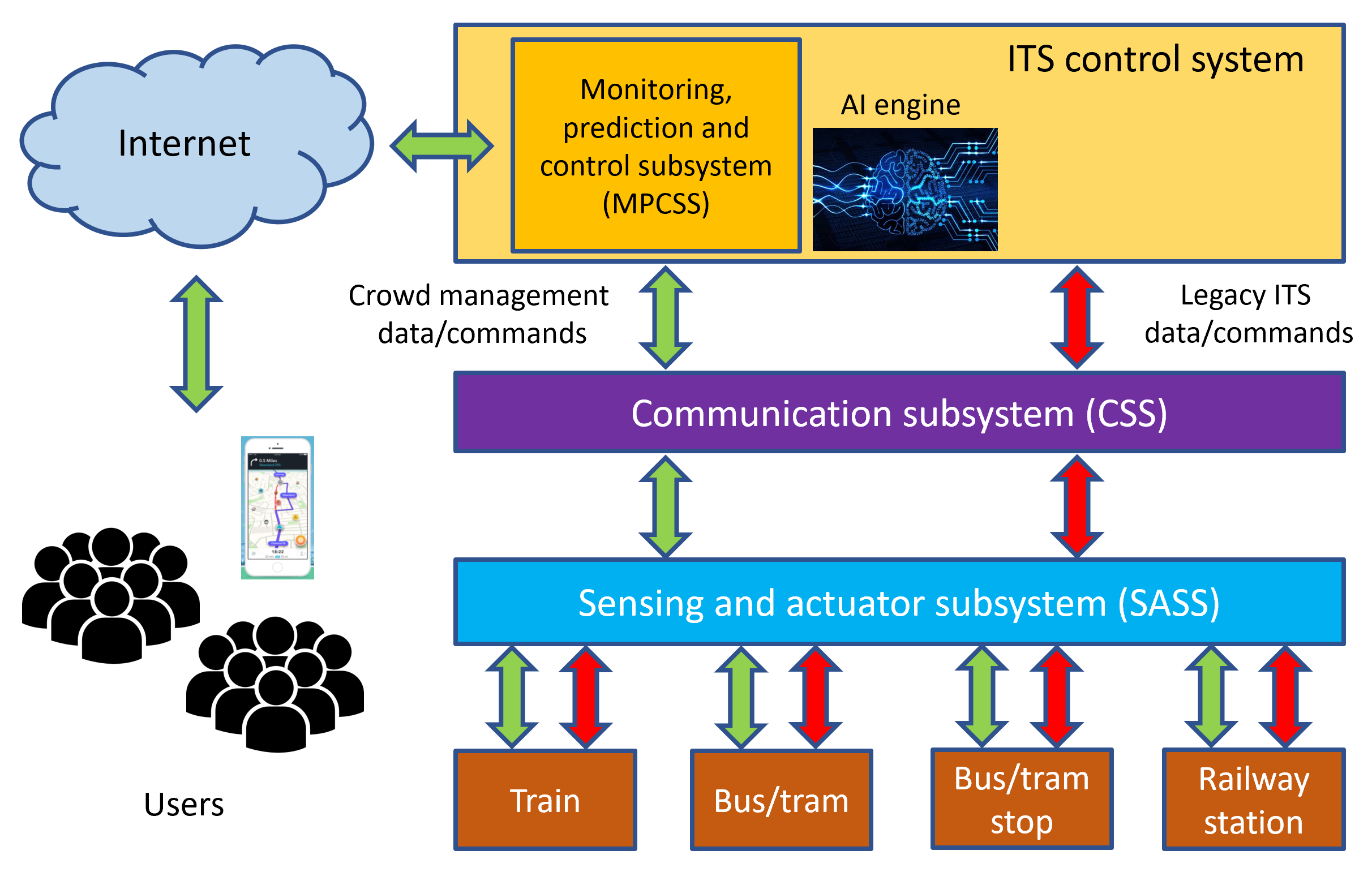}}
\caption{A crowd management reference architecture for ITS.}
\label{fig:architecture}
\end{figure*}


The key idea is to integrate heterogenous
sensing and communication technologies,
in dependence on the operation scenario and the ICT
infrastructure available in the urban area
where the system must be implemented.
To achieve such a goal in practice,
strong interdisciplinary design skills are needed,
including transportation engineering, telecommunications, computer science,
electronics, data analysis, and AI.

In our vision, the reference architecture
encompasses (Fig.~\ref{fig:architecture})
three subsystems:

\begin{enumerate}
\item the \emph{sensing and actuator subsystem} (SASS);
\item the \emph{communication subsystem} (CSS);
\item the \emph{monitoring, prediction and control subsystem} (MPCSS).
\end{enumerate}

The architecture involves
new data flows (marked as green arrows in Fig.~\ref{fig:architecture})
for crowd management,
in addition to existing data exchanges
(red arrows) commonly used in state-of-the-art
ITS systems for PT services.
The core and most innovative part of the system is the
SASS, with particular reference
to sensors for crowd monitoring:
in Section~\ref{sec:SASS}, we provide a
taxonomy and discussion of the
different sensing technologies
than can be utilized to this aim.
The actuator component mainly encompasses
two crowd-related flows of information:
(i) toward the users, carried out
by audio speakers, variable-message panels, or displays,
which are typically already present
in the PT system, or can be
readily installed at the bus/tram stops
as well as in railway/metro stations,
and also inside vehicles;
(ii) toward the PT operators (drivers and staff),
e.g., by means of evolved driver displays
that communicate to the drivers
decisions regarding PT adjustments
(e.g., holding, stop-skipping, or even route changing)
to be carried out in real-time.
It is worthwhile to note that,
with the advent of
\textit{autonomous-vehicle PT systems} \cite{Lam2016}
(based on driverless buses or metro trains),
the actuators can
directly modify
the vehicle behavior, without the need for human
intervention.

As regards the CSS, its characteristics
are strongly dependent
on the communications infrastructure
available in the urban area.
In general, this subsystem might encompass public wireless
networks (such as cellular networks)
and/or private wired and/or wireless
networks owned by the operator, such as, e.g.,
Global System for Mobile Railway (GSM-R)
or Long Term Evolution for Railway (LTE-R).
To cope with this heterogeneity,
it is envisioned that, at the protocol level stack,
the CSS can be readily interfaced
with the other subsystems by means of
standard or open interfaces
and/or using simple adaptation layers.

The MPCSS performs data collection and
real-time crowd prediction, possibly employing
AI and ML/DL techniques.
Based on such predictions, modifications
to the transport services can be implemented in real-time
(e.g., vehicle holding, stop-skipping,
overtaking, limited boarding, speed changing, short turning)
as well as at the strategical/tactical planning level
(e.g., optimization of timetables and routes).
The related control data are sent to
service operators (drivers, supervisors, etc.),
whereas service information, including RTCI,
is sent to the passengers by means
of displays and/or mobile transport apps.
This information could be notified by the same applications
to all the users of the PT, so as to discourage the access
to overcrowded stations and/or bus/tram stops,
and propose alternative travel solutions.
The MPCSS can be strongly integrated (and typically colocated)
with the ITS control system of the PT service
operator.

\section{Sensing technologies for crowd management}
\label{sec:SASS}

From the transportation perspective,
our taxonomy of sensing technologies
for PT systems considers (see also Fig.~\ref{fig:architecture})
the following
scenarios:

\begin{enumerate}
\item train and bus/tram vehicles (indoor scenarios);
\item railway/metro stations (indoor/outdoor scenarios);
\item bus/tram stops (mainly outdoor scenarios).
\end{enumerate}

The sensing solutions to be adopted
in these scenarios
belong to the
\emph{crowd analysis/monitoring} \cite{Zhan2008} family.
Detecting and estimating some features of a ``crowd''
in indoor and outdoor locations is an active research topic,
with many applications, including surveillance and security,
situational awareness, and emergencies
(for a recent review, see \cite{BendaliBraham2021} and references therein).
In PT applications, the feature of interest
is the number of components of the crowd
(i.e., \textit{crowd counting}) and/or
its density (i.e., \textit{crowd density estimation});
moreover, tracking of individuals in a crowd
could be needed to build O-D matrices, useful for
long-term planning.
A review of crowd analysis techniques
for urban applications,
including transportation systems, is provided in
\cite{Kaiser2018},
where approaches based on different
data sources, including information
fusion techniques,
are discussed and compared.

Sensing technologies for crowd analysis
can be classified \cite{Irfan2016}
as \emph{visual-based} (VB) solutions,
based on still or moving images/videos mainly acquired
by optical, thermal, or laser cameras,
or \emph{non-visual based} (NVB) ones,
which do not rely on images to estimate
crowd parameters,
but resort to other physical
quantities or features that can be related
to crowd parameters, such as, e.g.,
those of radio signals, temperature, or sound.

To perform crowd analysis,
VB techniques resort
to sophisticated image processing,
pattern recognition,
or computer vision techniques
\cite{Zhan2008,SilveiraJacquesJunior2010,Li2015}.
Indeed, thanks to recent advances in AI,
traditional camera sensors
are becoming ``smart'' and can detect,
recognize, and even identify persons.
VB technologies perform \emph{passive} sensing,
relying on a network of dedicated sensors,
without requiring active \emph{cooperation/participation}
of the users.
However, VB data are subject to stringent protection
regulations enforced by  international data privacy laws,
such as, e.g., general data protection regulation (GDPR) \cite{GDPR2016}
in the European Union.

Among NVB technologies (see \cite{Kouyoumdjieva2020}
for a recent review),
sensing solutions
based on mobile RF devices
represent an interesting approach,
due to the diffusion of smartphones and
other portable/wearable devices,
such as, e.g., pedometers, smart watches,
or biometrical sensors.
This approach to sensing is known as
\textit{mobile sensing} \cite{Ganti2011},
\textit{opportunistic sensing} \cite{Lane2010}
or \textit{participatory sensing} \cite{Burke2006}.
Data collected by means of
such \emph{device-aided} NVB systems
can be used not only
to count people in a crowd, but also to
gather additional information about individuals
(e.g., planned routes, O-D flows,
passengers using off-peak hours group ticket,
and so on).
However, a problem inherent to device-aided systems
is that they usually require
user cooperation/participation.
Another important aspect of NVB systems is that
they potentially collect sensitive data pertaining
to individuals, such as, e.g., daily movements
as well as home and work locations.
To motivate participation,
it is sometimes needed to introduce
incentive or reward mechanisms, or apply radical modifications
to the procedures to access
the PT service, such as an authentication phase
to use the service.
Although this issue could enhance
the aforementioned privacy concerns,
it could also be useful
as a means to increase the overall safety
of the PT systems during a pandemic outbreak,
by reducing the risk that
infected people can access the system.

When device-aided NVB techniques cannot be used
for crowd characterization, due to, e.g.,
lack of user cooperation and/or security/privacy issues,
RF-based \emph{non-device aided} or \emph{device-free}
approaches can be pursued, which operate by
analyzing the propagation channel variations
of wireless signals
induced by the people present in a given spatial area.
In limited cases, NVB sensing techniques
relying on physical properties
different from those of RF signals
(such as audio or ultrasound
signals) can also be employed.
From the user point of view,
information collected through device-free solutions is
less critical, as it may not affect users' privacy.

The main sensing
technologies for crowd management in PT systems
are summarized
in Table~\ref{tab:SAL-technologies}
and will be discussed
in the forthcoming
subsections. Table~\ref{tab:SAL-technologies} also
provides a preliminary classification of sensing technologies
on the basis of their degree of privacy. We defer to
Section~\ref{sec:chall} for a more detailed discussion about
privacy issues.

\begin{table*}[t!]
\caption{A taxonomy of sensing
technologies for crowd management
in PT systems.}
\defaultaddspace 4pt
\centering
\label{tab:SAL-technologies}
 \begin{tabular}{p{1mm}p{13mm}p{18mm}p{9mm}p{15mm}p{8mm}p{10mm}p{13mm}p{09mm}p{9mm}p{26mm}}
 \toprule
 & {\bf Technology}   & {\bf Frequency}
 & {\bf Max \newline range} &  {\bf Power \newline consumpt.~$^{\mathrm{a}}$}
 & {\bf User \newline cooper.} & {\bf Accuracy} & {\bf Processing \newline complexity}
 & {\bf Cost} & {\bf Privacy issues} & {\bf Scenario} \\
 \midrule
\multirow{3}{*}{\rotatebox{90}{\bf Visual-based\hspace{6mm}}}
 & Optical & Visible & $100$ m & Low & No & High & High & High & {Critical} & All \\
 \addlinespace
 & Thermal & {Infrared} & {$500$ m} & {Low} & {No} & {High} & {High} & {High} & {Moderate} & {Railway/metro station, \newline bus/tram stop} \\ \addlinespace
 & {LiDAR} &  {Ultraviolet, \newline Visible, \newline Near infrared}
 & {$1$ km} & {Medium} & {No}  & {High} & {Medium-high} & {Medium} & {Moderate} & {Railway/metro station, \newline bus/tram stop} \\ \addlinespace
 & {SAR}      & {Ku/Ka band} & {$10$ m} & {Very high} & {No} & {High} & {High} & {High} & {Moderate} & {Bus/tram stop}\\
 \addlinespace
  \midrule
\multirow{13}{*}{\rotatebox{90}{\bf Non-visual based\hspace{30mm}}}
 & {IR/PIR}   & {Infrared} & {$10$ m} &
 {Very low}   & {No} & {High} & {Low} & {Medium} & {Low} & {Bus/tram} \\ \addlinespace
 & Pressure   & {Mechanical} & {Contact} &
 {Very low}   & {No} & {High} & {Low} & {Medium} & {Low} & {Train and bus/tram} \\ \addlinespace
 & Acoustic   & {$0.01$--$100$ kHz} & {$20$ m}   & {Low} & {No} & {Low}  & {Low} & {Low} & {Moderate} & {Train and bus/tram, \newline railway/metro station} \\ \addlinespace
 & {IR-UWB}   & {$3.1$--$10.6$ GHz} & {$100$ m} & {Low}  & {No}  & {High} & {Low}  & {Medium} & {Low} & {Railway/metro station, bus/tram stop} \\ \addlinespace
 & {RSSI/CSI} & {Various}  & {Variable} & {Very low} & {No}  & {Medium} & {High} & {Low} & {Low} & {Railway/metro station, bus/tram stop} \\ \addlinespace
 &  {RFID}    & {$13.56$ MHz} & {$10$ cm} &
 {Very low}   &  {Yes} & {High} & {Low} & {Medium} & {Critical} & {Train and bus/tram, \newline railway/metro station}  \\ \addlinespace
 &  {NFC}     & {$13.56$ MHz} & {$10$ cm} &
 {Very low}   &  {Yes} & {High} & {Low} & {Medium} & {Critical} & {Train and bus/tram, \newline railway/metro station}  \\ \addlinespace
 & {Bluetooth}      & {$2.4$ GHz} & {$50$ m} &
 {Very low}   & {Yes} & {Medium}  & {Low} & {Low} & {Critical} & {Railway/metro station, \newline bus/tram stop} \\ \addlinespace
 & {WiFi}      & {$2.4/5$ GHz} & {$100$ m} & {Medium} & {Yes}
 & Low & {Medium} & {Low} & {Critical} & {Railway/metro station, \newline bus/tram stop}\\ \addlinespace
 & {LTE} & {$800/1800/2600$ MHz} &
 {$10$ km} & {High} & {Yes}  & {Low} & {High} & {Medium} & {Moderate} & {Railway/metro station, \newline bus/tram stop} \\ \addlinespace
 & {5G}  &  {Sub-6 GHz} & {$10$ km} & {High} &
 {Yes} & {Low} & {Low}  & {Medium} & {Moderate} & {Railway/metro station, \newline bus/tram stop} \\ \addlinespace
 & {5G}  &  {MMW band} & {$1$ km} & {High} &
{Yes} & {Medium} & {Low} & {Medium} & {Moderate} & {Railway/metro station, \newline bus/tram stop} \\
\addlinespace
 \bottomrule
\multicolumn{10}{p{100mm}}{$^{\mathrm{a}}$ Power consumption refers to a single sensor.}
\end{tabular}
\end{table*}


\subsection{Infrared sensors}

Infrared sensors (IR) are commonly
used in traditional APC
systems for counting the number of passengers
boarding or alighting a vehicle (usually a bus or a tram).
Commercially-available solutions
employ a couple of IR sensors
(acting as a transmitter/receiver)
forming a ``light barrier'' aimed
at detecting the passage
of people at the input/output gates
of the vehicle.
In alternative, a single \emph{pyrolectric} IR (PIR)
sensor can be used, which detects
the IR radiation emitted by the human body
in the wavelength range $2$--$14$ $\mu$m.
To enhance reliability, combined installation
of IR/PIR sensors
can also be conceived.
Although IR sensors are very
frequently used in PT systems, with counting accuracy generally
well above $90\%$ \cite{Olivo2019},
their performances worsen  when multiple people board
through the same door simultaneously
\cite{Chen2008,Choi2018}.
Moreover, installation
could be expensive, since typically
more than one barrier
per door is needed to detect the passenger flow.

\subsection{Pressure/load sensors}
\label{sec:load-sensors}

Another solution commonly
employed in traditional
APC systems counts the
number of passengers boarding or
alighting buses or trams
by means of pressure-sensitive
switches (``treadle mats'')
placed on the vehicle steps,
which are activated under the effect
of the passenger weight.
These solutions are simple, accurate (above $95\%$) and rugged,
ensuring long operational life:
installation of multiple treadle mats
over different steps allows one to
discriminate between infeed/outfeed
motion \cite{Pinna2010}.
An innovative pressure sensor is the
Velostat/Linqstat one,
which is a carbon-impregnated conductive
polymeric foil
that can be used as a low-power
inexpensive pressure sensor. Such a sensor
has been adopted in
\cite{Vidyasagaran2017} to implement
a system aimed at monitoring
seat occupancy in a bus.

Another application of pressure/load sensors,
placed on the ground or on the suspensions,
is in \emph{weigh-in-motion} (WIM) systems,
which estimate the number of passengers by
the loading of the vehicle detected
before and after the stops.
Since most of the modern trains are equipped
with electronic weighing sensors
providing information to the braking system,
a WIM solution exploiting such sensors
has been proposed and
implemented in the Copenaghen metro system
\cite{Nielsen2014}.
In \cite{Kotz2016} two algorithms
for passenger counting
are proposed, which estimate the passenger
load on the basis
of the pressure variations
of vehicle air ride suspensions,
which are commonly employed
in almost all modern transit buses.
WIM systems represent a convenient solution
to measure crowdedness inside vehicles, even though
it can be difficult to infer
the actual number of passengers
boarding or alighting \cite{Pinna2010}.

\subsection{Optical cameras}

Optical cameras are widely used in
private and public spaces for surveillance
and security, like in closed
circuit television (CCTV) systems,
and are routinely installed inside PT vehicles and
stations to this aim.
Since often they can support
crowd monitoring functions
with firmware/software upgrades,
optical cameras are among the most versatile
and used techniques for crowd analysis
\cite{SilveiraJacquesJunior2010,Chen2013}.
Most works on crowd counting and detection
based on optical cameras rely
on computer-vision technologies
\cite{Chen2008,Chato2018,Zhang2021},
wherein crowds are detected using specific features,
e.g., facial recognition or motion
tracking, extracted from
images/videos.
Moreover, the majority of
recent works employ AI or ML/DL
algorithms \cite{Abdou2020}.

Camera-based solutions can be applied in all transportation scenarios,
both indoor and outdoor ones.
Video cameras mounted
inside road or rail vehicles can be
used to estimate the number of passengers
and their flow (i.e., whether
they are boarding or alighting)
\cite{Yu2007,Chen2008,Yang2010,Sun2019}.
An image-processing technique based on a modified
Hough transform
is proposed in \cite{Yu2007},
aimed at detecting the contour features of
heads and estimating accordingly
the number of passengers
and their flow in a bus.
Many recent approaches
are based on CNNs, such
as the passenger counting system
proposed in \cite{Liu2017}, which
also exploits the spatio-temporal properties
of video sequences acquired on a PT bus
in China, or the solution proposed in \cite{Meghana2020},
where crowd density inside a bus is detected
and classified
in $5$ different levels (from very low to very high),
to be displayed
by LCD screens installed
at the bus stops.
A deeply-recursive CNN-based solution is proposed
in \cite{Ding2018} and tested on a dataset of
images taken at the Bus
Rapid Transit (BRT) in Bejing, China.
A neural-network based crowd density
estimation algorithm
is described in \cite{Cho1999},
targeted at underground
station platforms,
which has been experimentally
tested on a sequence of CCTV images
acquired at a metro station in Hong Kong.
Finally, VB techniques for counting people
at bus stops are proposed
in \cite{GarciaBunster2009,GarciaBunster2012},
based  on computer-vision techniques,
which process measurements of foreground areas
corrected by suitable perspective
transformations.

In summary, crowd analysis based on optical sensors
is versatile and powerful, but has several limitations:
camera sensors, indeed, are expensive,
and each camera can only cover a small area,
resulting in high deployment costs in complex
and/or large environments.
In some scenarios, optical camera-based systems do not allow to estimate
the number of people with sufficient accuracy,
due to possible obstructions, clutter, and
poor light/weather conditions.
Moreover, crowd analysis techniques
based on images require a high
computing power.
Finally, management of optical
camera-based systems might be
cumbersome, due to privacy-related
restrictions.


\subsection{Thermal cameras}

Thermal cameras can detect people in low-light environments,
complete darkness, or other challenging conditions,
such as smoke-filled and dusty environments \cite{Andersson2009}.
Thermal imaging cameras are currently used in some countries
to prevent accidents and infrastructure damage in PT systems,
detecting, for example, people
walking on the tracks or fire events.
This technology, however, can be employed also to monitor
crowding situations both in indoor and outdoor
scenarios.

Some people counters based on thermal cameras
are proposed in \cite{Lin2011,Tikkanen2014,Tyndall2016}.
Different from optical cameras, people counters
based on thermal cameras are
less sensitive to the level of ambient lighting
or background colour contrasts, but their
performance can be adversely affected
by heat sources and weather conditions.
In addition,
real-time image processing
can be computationally intensive.
The high costs of instrumentation still limit
the widespread use of thermography for
crowd monitoring.

Recently, hybrid approaches,
combining thermal and optical imaging sensors,
and intelligent processing
(based on DL and CNNs),
have been implemented to improve the accuracy
and real-time processing of camera-based systems \cite{Amin2008,Yaseen2013}.
In \cite{Gomez2018}, a people
counting algorithm is developed, which uses
low-resolution thermal images and employs small-size CNNs,
being able to run on
a limited-memory and low-power platform.

\subsection{LiDAR}

Another option to detect and track persons is represented by
light detection and ranging (LiDAR) sensors, especially
in environments where there are several interacting people
\cite{Jiang2021}.
A LiDAR sensor is a distance-measuring system that works by illuminating
the target with a laser beam and sensing the reflected laser light.
There are different implementations of
LiDAR sensors based on their coverage
area and wavelengths.

Compared to traditional VB approaches,
LiDAR is less sensitive to varying lighting conditions and
requires, in general, lower data-processing times.
Thanks to these features, LiDAR-based counting systems are suitable
to all transportation scenarios, both indoor and outdoor.

A real-time 2D LiDAR monitoring system
for people counting is proposed in
\cite{Lesani2020}, which turns out to be useful
in monitoring wide areas and dense groups of persons.
The solution proposed in \cite{Hashimoto2015}, instead,
employs two LiDAR sensors set at different heights,
aimed at detecting people's heads and knees,
to improve tracking performance.
To increase system accuracy, other solutions
use three-dimensional (3D) LiDARs:
in \cite{Akamatsu2014}, for example,
a technique for people counting is presented,
which works well even if two or three persons
pass at the same time.
However, an important limitation of 3D LiDAR is
represented by its computational cost, which can be even higher
than optical camera-based solutions.

\subsection{Acoustic/ultrasound sensors}

Acoustic sensor-based approaches perform people counting by relying on
audio signals transmitted by smartphones or produced by speaking people \cite{Kouyoumdjieva2020}.
A crowd counting solution based
on audio tones is presented in \cite{Kannan2012}, leveraging
the microphones and speaker phones available
in most mobile phones.
Effectiveness is proven through
several experiments at bus stops or aboard, which show, however,
that counting latency can significantly grow as the number of devices
increases; as a consequence, this technique may be appropriate only for
low-density scenarios.

Despite their simplicity, applicability of purely
acoustic-based solutions in crowd counting
is largely limited by their high sensitivity
to environmental noise level.
For this reason,
hybrid solutions, employing not
only acoustic sensors, have been proposed.
In \cite{Agarwal2015}, for example, a multimodal
sensor network is built, which exploits sound intensity in conjunction
with additional information sources,
like carbon-dioxide level and
RF link strength,
in order to increase estimation accuracy.
Another hybrid solution is presented
in \cite{AlHafizKhan2015},
which proposes an opportunistic collaborative
sensing system, based on acoustic and motion
sensors integrated in smartphones.

A different option to count people
is represented by acoustic-based solutions
using ultrasound sensors, which
perform well for indoor spaces and small crowds
\cite{Kouyoumdjieva2020}.
When the reverberation of the transmitted waves
is received, the number of people can be estimated by exploiting
information from the receive time or the signal decay.
Based on this approach,
an ultrasonic sensing technique
for estimating the number of people
is presented in \cite{Shih2015},
which exhibits satisfactory performance
when the occupancy of the space
does not reach its maximum.


\subsection{Device-free RF sensing}
\label{sec:device-free}

Device-free crowd analysis based
on RF signals is an emerging
technique,  which does not require
installation of expensive cameras
nor it suffers from privacy related concerns.
It exploits the impact
of the monitored
crowd on RF communications
to infer information
regarding the size and/or density of the crowd,
either  using traditional
radar methodologies (range and Doppler
analysis) \cite{DiDomenico2016,Bartoletti2017,Brena2021},
mainly with impulse-radio (IR) ultrawideband (UWB) signals \cite{Choi2018,Choi2021},
or by analyzing features extracted by
channel quality measurements,
such as received signal strength  indicator (RSSI)
\cite{Yuan2013,Depatla2015,Ibrahim2019}
or channel state information
(CSI) \cite{Xi2014,DeSanctis2019,Oshiga2019}.

People counting using RF signals can be used in dimly lit
places, and in smoky and dusty environments:
hence, it represents an interesting solution
for metro/subway stations and/or at bus/tram stops.
In \cite{Choi2018} a people counting
algorithm using a couple of IR-UWB
radar sensors is proposed,
which was tested in a subway
platform in Seoul, South Korea,
showing accuracy values higher than
$90\%$.
A solution to count the people
in a queue, based on RSSI measurements carried out by
Bluetooth low-energy (BLE)%
\footnote{BLE is a low-energy consumption
version of Bluetooth standard, which assures
better communication
performances with a limited power consumption.}
devices, is described in \cite{Brockmann2018}:
the system is completely passive and
estimates the number of persons in the queue
by analyzing the mean and variance of
the RSSI values between a BLE beacon and a receiver
covering a certain area.

In general, RSSI-based approaches for crowd analysis exhibit
good performances in small monitored environments, where
the propagation channel variations
are dominated by the attenuation caused by the
people actually present in the environment.
On the other hand, in a rich
scattering environment,
crowd analysis approaches based on CSI
provide a more reliable people counting,
but are considerably more complex.
Belonging to this class, in \cite{Sobron2018} and
\cite{DeSanctis2019}
the received WiFi and LTE
downlink control signals
are processed, respectively,
to extract the changes of the
propagation channel induced by the presence of different
number of people: the related
counting algorithms
exhibit variable levels
of accuracy in different scenarios.

Finally, more sophisticated
microwave SAR tomography
techniques \cite{Reigber2006,Capozzoli2010}
can also be used, which provide specific RF
images from which
more detailed crowd information can be extracted,
by using complex image classification algorithms
(i.e., count, distribution, mobility, etc.).
Modeling issues induced by the intrinsic
near-field scenario (e.g., typical for bus/tram stops)
could be overcome using
specialized algorithms \cite{Capozzoli2011}.

In summary,
excluding SAR-based techniques,
device-free RF sensing
is a moderate-complexity crowd analysis
technique with
low installation costs.
Moreover, RF signals can
penetrate obstacles to a certain extent
and are not affected by weather/illumination.
A moderate accuracy (around $80\%$ for RSSI-based techniques,
up to $90\%$ and larger for CSI-based ones) can be expected
in simple scenarios,
but it is questionable whether this approach can be
scalable to large crowds,
especially in complex propagation
scenarios \cite{Kouyoumdjieva2020}.
Moreover, counting algorithms
based on this approach
usually require a site-specific training phase,
which complicates practical
installation and maintenance.


\subsection{Device-aided RF sensing}
\label{sec:device-aided}

In device-aided approaches
for crowd analysis, users are expected
to carry RF devices, which must be switched on
to enable people counting.
Since modern smartphones
are ubiquitous and, moreover, are equipped
with several sensors and multiple RF interfaces,
device-aided solutions
are commonly used to gather different types
of information for many different purposes
and applications (see \cite{Kouyoumdjieva2020}
for a recent review).

\smallskip
\subsubsection{RFID and NFC}

RFID-based solutions require that the passengers carry
passive RFID tags, whose presence
can be detected by a reader.
The Authors in \cite{Oberli2010}
propose an APC system for bus vehicles
employing commercial
EPC Gen2 tags, which are recognized by a reader
located in correspondence of the bus gate.
A similar short-range communication system
is Near Field Communication (NFC)
\cite{Want2011} technology, which is currently
supported by many modern
smartphones and tablets.
When a device with NFC functionalities
appears in the reader's working range,
which can be placed at the station gates
or at any other fixed access point,
it ``wakes up''  and sends a signal
containing encoded data.
Finally, it should be observed
that RFID and NFC technologies
are used in traditional and emerging AFC
systems for electronic ticketing, such as
MIFARE contactless cards
or mobile-based payment systems
\cite{DemirAlan2018}, which can also
be employed for passenger counting.
For these applications, privacy is characterized by the ability of unauthorized
users to trace RFID and NFC devices using their responses
to readers'  interrogations.
Since RFID and NFC devices are typically not tamper-resistant, an adversary
can capture them and expose their secret parameters.

\smallskip
\subsubsection{Bluetooth}

Bluetooth is a consolidated short-range
RF technology, supported by almost all
smartphones on the market:
crowd monitoring algorithms using Bluetooth have
been proposed in several papers
(see e.g. \cite{Weppner2014} and references therein).
An algorithm
based on off-the-shelf Bluetooth hardware
for counting bus passengers
has been proposed in \cite{Kostakos2010}
and tested in the city of Funchal, Portugal.
The system consists of a Bluetooth
scanner mounted on the bus ceiling,
which periodically
scans for discoverable Bluetooth
devices in its range,
and is aimed at discovering O-D relations
by post-processing data and correlating them
with information related to bus location
and tickets issued
by fare machines.

Bluetooth can also represent an
efficient solution for crowd counting at
bus/tram stops.
A crowd analysis solution based on
BLE is proposed in \cite{Basalamah2016},
where a large population carry
BLE proximity tags, acting as
beacons, whose presence
is sensed by smartphone carried
by few volunteers.
A reciprocal solution
can be used at bus stops, where
BLE beacons are installed
at the bus stations, and are detected
by passenger smartphones in close proximity
of the stops.%
\footnote{This solution relies on the same
technology introduced by Google and Apple
in the most recent versions of
their smartphone operating systems,
and is used by many national
contact-tracing apps (e.g., Immuni
for Italy \cite{Immuni}).}

Compared to WiFi-based sensing (discussed later),
Bluetooth devices are cheaper,
less power-hungry
and are characterized
by increased flexibility and simplified
installation.


\smallskip
\subsubsection{WiFi}

Many crowd analysis solutions exploit the
characteristics of the IEEE 802.11
(WiFi) protocol, which is widely
used by passengers during their trips.
The technique adopted
in \cite{Handte2014,Handte2016}
estimates the current number of passengers
in a bus  by counting the number of
\emph{probe requests}, i.e.,
medium access control (MAC) addresses,
sent by WiFi-equipped
devices in the vehicle.
A similar approach is followed in
\cite{Nitti2020}, where
a de-randomization
mechanism is introduced
to counteract software randomization
of MAC addresses,
recently introduced in many
operating systems.

One of the problem inherent to the use
of WiFi-based techniques for crowd counting inside vehicles
is the ability to distinguish
between people outside the vehicle
and actual passengers.
This issue was tackled in \cite{Handte2014,Handte2016}
by filtering the probes with a sliding window,
aimed at removing MAC addresses
that were not detected
over a longer period of time.
In addition, in \cite{Nitti2020}
the received power is also
used to discriminate devices
that are likely to be outside the bus.
The system in \cite{Handte2014,Handte2016}
was able to detect
only around $20\%$ of the passengers in a real setting, since
several users may have turned off
the WiFi interface.
{Underestimation of the number of
passengers is a common problem for these techniques,
which can be compensated by a proper calibration of the procedure
in each scenario of interest.

The main advantage of WiFi-based
techniques is that
they allow to track passengers also when they alight the bus,
allowing to estimate O-D flows.
WiFi-based counting can also
be employed in metro/railway stations,
since access points are typically available
in such scenarios.


\smallskip

\subsubsection{Cellular}

Similarly to WiFi,
cellular signals such as LTE and 5G ones
can be used for device-aided
crowd counting,
thanks to their ubiquitous
availability and good penetration
in indoor environments.
Cellular signals
could be available in
areas where the WiFi coverage
is not present, such as bus/tram stops,
remote and small railways/metro stations.
A cellular-based crowd density estimation method
is proposed in \cite{Shibata2019},
which measures the signal strength
emitted in uplink by the
smartphones of the crowd components,
and classifies the crowd density in
different levels using DL techniques,
with an accuracy of $78\%$ when three
levels are considered.

In principle, the position
of users in a cellular network
can be obtained with satisfactory
precision by combining knowledge
of the serving base station, RSSI values,
and triangulation principles \cite{PeralRosado2018},
which can be at the basis
for large-scale crowd analysis.
However, this approach is difficult to be used
in real-time, whereas
it is more suited for long-term
travel demand estimation \cite{Demissie2016}.
Real-time passenger counting
using cellular data
is rarely performed,
due to several drawbacks:
(i) it requires gathering data
from different mobile operators;
(ii) it raises significant privacy concerns;
(iii) it does not allow one to precisely
discriminate passengers from general public
in open areas.
A breakthrough could be
the planned introduction in 5G systems
of the millimeter-wave
(MMW) band, which will require
very small cells:
from the viewpoint of crowd analysis,
the placement of small cells
allows one to more precisely monitor
spatially limited areas,
like bus/tram stops.

\subsection{Discussion}
\label{sec:discussion}

In this Section, we provide a discussion
regarding the applications of the
above-mentioned sensing techniques in the different
transportation scenarios.%
\footnote{
For a review and discussion of some industrial solutions employing
some of these technologies for people
counting in PT systems, see \cite{Kouyoumdjieva2020}.
}


\smallskip

\subsubsection{Trains/buses/trams}

Such vehicular scenarios are characterized
by a well delimited indoor space with
a limited number of accesses
(gates).
Traditional APC systems
\cite{Pinna2010,Grgurevic2022,Nitti2020}
count the number of passengers
inside vehicles on the basis of
various onboard sensors,
mainly IR or pressure-sensitive ones.
The number of passengers
can also be estimated
by the number
of validated/issued
tickets, as in AFC systems, which
however requires user cooperation
and can provide underestimated
results in case of diffuse
fare evasion.

Although solutions based on
sensors installed on the vehicles
are simple, they require a large
initial investment. Therefore,
the diffusion of portable
and mobile devices between passengers
have pushed many researchers to study
solutions based on RF techniques
(both device-aided and device-free).
Summarizing, as also indicated
in Table~\ref{tab:SAL-technologies},
even though more sophisticated VB/NVB sensing
technologies may be used as well,
their usage is not expected to lead to significant
innovations in this scenario, compared to
traditional
solutions adopted in commercial
APC systems.


\smallskip

\subsubsection{Railway/metro stations}

Many crowd monitoring options are
available in this scenario, since
the access to the stations
occurs through a limited number of gates.
Morover, CCTV surveillance systems are typically present inside stations
and can be used for VB crowd analysis.
The access to train platforms
is usually governed by turnstiles
where tickets/passes must necessarily
be validated:
thus, IR-based APC or AFC systems could be a reasonable option
to count passengers in this scenario.
However, this solution
only counts people
trying to access the platforms,
disregarding other
people which could walk
inside the station for different
purposes (e.g., for shopping or leisure).
RF-based techniques could suffer from
coverage problems, especially
within metro/subway stations.


\smallskip

\subsubsection{Bus/tram stops}

This scenario is by far the most difficult to manage, since
bus/tram stops are usually located in
outdoor spaces, not well delimited by
fixed gates.
In this case, both VB and NVB
sensing technologies can be used,
but it is imperative to adopt
cost-effective, rugged, and low-power
solutions, in order to reduce
the maintenance cost.
Moreover,
solutions that do not require
significant infrastructures are preferred,
since in many cases the
stops are not equipped with shelters and are
indicated by simple poles.

\begin{table*}[!t]
\caption{Main advantages related to adoption of crowd management in PT systems.}
\label{tab:SAL-advantages}
\centering
\begin{tabular}{p{0.20\textwidth}p{0.50\textwidth}}
\toprule
{\bf Actor} & {\bf Advantages} \\
\midrule
PT service operator (planner, manager) & Having real-time crowding data of the different
segments of the PT system allows one to plan services more efficiently and
to quickly readapt them to tackle critical situations, localized in space and time \\
\addlinespace
PT operators (drivers, supervisors, staff) &
Knowing in advance
crowding situations at the stops or stations
allows one to tackle in real-time critical situations
(e.g., implementing holding or stop-skipping strategies,
or following alternative routes) \\
\addlinespace
PT users & Knowing crowding situations allows one to use
alternative transportation modes or
make the trip in another hours,
if not strictly necessary;
by the reservation system, if available, the users can access
vehicles without unnecessary crowding
at the stops or stations \\
\addlinespace
Police forces & Knowing in real-time and/or predicting possible
crowds - potentially dangerous for public health and/or for public order -
allows a more timely and targeted intervention \\
\addlinespace
Sanitary system & Smart reduction of crowding resulting from an agile
management of the PT system allows one to reduce the diffusion of
infection and prevent further outbreaks \\
\addlinespace
General population &
An improved QoE of the PT system
entails a reduction of private car usage and
of pollution in urban areas, with benefits on energy
consumption and climate change \\  \bottomrule
\end{tabular}
\end{table*}

\section{Main innovations and advantages}
\label{sec:i-a}

The crowd management functionalities
of the new framework
can provide several innovations and advantages
that are not present
in state-of-the-art ITSs:

\smallskip

\subsubsection{Proactive control of station access}

In railway/metro applications,
on the basis of the knowledge of
the number of passengers aboard the
arriving trains and the prediction of
those alighting at the station,
it will be possible to predict the number
of accesses to stations/platforms with a low error
margin and in real-time, so as to avoid crowding.
This number can be communicated to the users
(by displays at the station gates or by the
mobile transport apps) and can
be used by the security
operators to filter passengers at the turnstiles.
Priority policies can be envisioned, such as
taking into account the time already spent in queue,
or the trip motivations (e.g., a priority
could be assured to health workers,
disabled or elder users,
law enforcement operators, and teachers/students).

\smallskip

\subsubsection{Vehicle access reservation}

In bus/tram trips, a vehicle access reservation
system can be implemented. A sensor at the
bus/tram stop detects the user presence
and exchanges information with
his/her device (i.e., the smartphone),
so as to grant him/her the access to
board the first arriving vehicle
(a virtual queuing system) or
putting him/her in an \emph{overbooking list}
(with priority) to allow him/her to board the
next one. The application can generate an e-ticket
with the access grant (e.g., a QR-code)
that can be validated on board at the ticket
machine.

\smallskip

\subsubsection{Crowding information dissemination}

Users receive real-time
information related to available
capacity (in terms of number of seats or in percentage)
of the bus/trams/trains in arrival and/or
crowding at the stops/stations,
so as to avoid unnecessary
waiting or crowding, and possibly
reschedule their trips.
Such RTCI can be provided
by means of displays installed in correspondence of
the stops or at the entrance
of the stations, or by messages/alerts
issued by mobile transport apps.

\smallskip

\subsubsection{Crowd-based route planning}

Users can plan their trips on the
mobile transport app, by taking into
account not only geographical
information and traveling times (static data),
but also traffic and crowding information
about vehicles and stops/stations
during the trip (dynamic data).
The app may not necessarily suggest the
shortest route, but the least crowded one, taking
into account also crowding levels
measured during the trip.
Such a feature not only helps
reduce crowds (and the consequent
infection risk in case of a pandemic), but
also improves the passenger QoE, by
distributing more efficiently the load
on the transportation network.

\smallskip
\subsubsection{Crowd-aware real-time control}

Some of the typical functions of
PT planning and operations can benefit
from the availability of the new crowd information.
At the strategic and tactical levels,
average long-term  crowding data can
be used to plan the services. At the real-time and operational level,
critical situations (service disruptions, unusual crowding) or even random
spatio-temporal demand variations
can be tackled by implementing crowd-aware
rescheduling solutions, such as holding,
stop-skipping, or similar ones,
which can be communicated to the drivers and
staff in real-time, as well as to the passengers
via displays and/or alerts.

\smallskip

The main advantages of the crowd management framework
are summarized in Table~\ref{tab:SAL-advantages}.
Compared to static methods, like traditional
survey-based compilation
of O-D matrix flows,
more efficient planning
and real-time control of the operation of the PT system
is allowed.
The large amount of generated data can be used by AI and ML/DL
algorithms to better understand
and plan a series of aspects
generally associated with improvements of the
quality of life in urban areas and smart cities.

Compared to the other anti-COVID-19 solutions discussed
in Section~\ref{sec:mobility},
the new framework is not aimed at
enforcing social distancing measures only.
Its scope is wider, since it tries to incorporate crowding information
to adaptively optimize the overall performance of the PT system.
As a by-product, it  also allows to partially recover the drawbacks and
inefficiencies of PT systems due to the adoption of  rigid social distancing
measures during both pandemic and post-pandemic phases.



\renewcommand{\arraystretch}{2}
\setlength\arrayrulewidth{1pt}
\begin{table*}
\centering
\caption{A colortable showing the
impact of the main challenges on the subsystems of the
introduced reference architecture: the intensity of the color
varies from darker (mapping an higher impact) to lighter
(mapping a lower impact).
}
\label{tab:challenges}
\setlength\arrayrulewidth{1pt}
\begin{tabular}{| m{2.5cm} | >{\centering\arraybackslash}m{1.1cm}| >{\centering\arraybackslash}m{1.1cm} | >{\centering\arraybackslash}m{1.1cm} |}
\hline
\textbf{Challenge} & \textbf{SASS} & \textbf{CSS} & \textbf{MPCSS} \\
\hline
Cooperative sensing & & \cellcolor{blue!40} & \cellcolor{blue!40} \\ 
\hline
Data fusion & & & \cellcolor{blue!100} \\
\hline
Security & \cellcolor{blue!10} & \cellcolor{blue!10} & \cellcolor{blue!10} \\ 
\hline
Privacy & \cellcolor{blue!40} & \cellcolor{blue!40} & \\
\hline
Communications & & \cellcolor{blue!100} & \\
\hline
Crowd control & & & \cellcolor{blue!100} \\
\hline
\end{tabular}
\end{table*}
\renewcommand{\arraystretch}{1}


\section{Main challenges}
\label{sec:chall}

In this Section, in relation to
the proposed  reference architecture,
we enlighten some of the main challenges
to be addressed
for the introduction of crowd management
functionalities in existing or forthcoming ITS systems.
The impact of such challenges on the subsystems of
the introduced reference architecture is summarized
in Table~\ref{tab:challenges}.

\smallskip

\subsubsection{Cooperative sensing}

In NVB cooperative sensing approaches,
a large number of users
must be engaged to obtain significant
amounts of data.
Several works \cite{Zhang2016, Jaimes2015}
have proposed to introduce
incentive mechanisms aimed at stimulating
user motivation and
encouraging participation.
A common solution is
to let participants earn credits in exchange
of their data.
However, such a strategy might
present some privacy risks,
by exposing sensitive data and linking
them to users' identities.
Several efforts have been made
to propose privacy-preserving cooperative sensing
approaches (based, for example, on data anonymization,
randomisation, and aggregation).
An interesting solution
is discussed in \cite{Klopfenstein2018},
where a rewarding platform
based on a voucher exchange system is proposed,
which decouples mobile
crowd sensing instruments from participation incentives.
Each voucher is produced as a compensation
for user's participation, and it is designed to
be fully anonymous and not exclusive.


\smallskip
\subsubsection{Security}

As an ICT-based evolution of the state-of-the-art ITSs,
the proposed reference architecture strictly relies
on ICT infrastructures too, i.e.,
most of the functions accomplished by
the MPCSS subsystem could be implemented using modern
cloud technologies.
Since every ICT infrastructure
can be prone to well-known
cyber-attacks, security issues are particularly serious
in all portions of our reference architecture, mainly due to
the large volume of data exchanged among subsystems.
A malicious user, for example, could launch a man-in-the-middle attack to the system,
by intercepting and altering the content of the exchanged
messages between crowd monitoring tools and vehicles,
which could entail catastrophic effects
in the decision-making phase, especially when
self-driving vehicles are involved.
Furthermore, denial-of-service attacks could be used
to saturate the resources of the infrastructure
to completely interrupt the decision-making process.}

Security issues related to the IoT paradigm are well-known
in the literature, but there are some aspects peculiar
to ITSs \cite{Hahn2021}.
In particular, one of the most limiting factor
is represented by the scarce computational
resources of many sensing devices used in ITSs, which may render
the commonly used countermeasures
(i.e., cryptography and authentication schemes)
infeasible for providing secure communications.
The proposed architecture, indeed, is based on
a plethora of heterogeneous sensors and devices,
which are required to be inexpensive, low-energy, and, in some cases,
have small form-factor.
However, energy-constrained sensors are resource-limited
in terms of memory, computational capabilities, and communication range.
Such constraints greatly limit the use of complex algorithms,
useful even to guarantee security and privacy.
Therefore, it becomes paramount to adopt
energy-aware solutions to prolong the lifetime
of the sensing subsystem.
In this sense, a promising solution
can be ambient backscattering \cite{Darsena2019},
where small passive devices are able to transmit data
by reflecting electromagnetic waves transmitted
by existing RF transmitters.


\smallskip
\subsubsection{Data fusion}

The availability of large amounts
of data acquired by heterogeneous sensors can pose
a challenging problem for crowd monitoring, prediction,
and control tools.
In the proposed architecture, indeed, the MPCSS subsytem must
analyze in real-time a large amount of multi-source
and multi-modal data, which are
characterized by different levels of resolution, accuracy,
reliability, and redundancy.
Various data fusion (DF) algorithms,
aimed at associating, correlating, and combining information
from multiple sensors, are commonly adopted
to provide accurate and
timely decision-making support.
Among the commonly adopted approaches
(i.e., statistical, probabilistic, and data-driven ones \cite{Li2021}),
probabilistic-based methods seem to be more
suitable in our scenario.
In particular, Bayesian approaches,
maximum likelihood methods, and Kalman filter-based DF techniques
can be utilised for
multi-sensor data fusion
\cite{Faouzi2011}.

\smallskip
\subsubsection{Privacy}

Privacy is one of the major issues related to
sensing technologies for crowd management. In what follows,
VB and NVB techniques are discussed separately,
since they involve different privacy concerns that can be
overcome by resorting to radically
different technical solutions.

VB solutions must be designed to implement
data protection principles, collectively referred to as \textit{privacy-by-design},
set out in international data privacy laws.
For crowd counting and crowd density estimation,
identification is not necessary. Therefore, for such applications,
privacy-by-design can be achieved by suitably choosing resolution
and other modifiable factors of VB devices to ensure that no
recognizable facial images are captured.
In addition to privacy-by-design,
some data privacy laws also mandate \textit{privacy-by-default}.
Under such an obligation, personal data collected through VB approaches
must be used only for the specific purpose of crowd management.
This means that the minimum required amount of personal data
should be collected, their processing should be minimized,
and their storage and accessibility should be controlled.
A viable approach to ensure privacy-by-default consists of turning
VB feeds featuring individuals into numbers and heatmaps
in such a manner that the data subject is not or no longer identifiable.
Within the European Community, such countermeasures
allow VB solutions to comply with the GDPR, which is one of the most restrictive
legislations in the world in terms of safeguarding citizens' privacy.

Regarding NVB solutions, sensing solutions
based on mobile RF devices pose important privacy issues, especially
for device-aided crowd monitoring.
Apart from the issues regarding
participatory sensing, which
have been discussed previously,
many device-aided RF-based NVB techniques
basically perform monitoring of over-the-air
beacon signals sent by individuals' smartphones
when they connect to a network or install an application.
In this case, to extract relevant features for crowding information,
processing of devices' IDs is required, e.g., MAC addresses in WiFi networks.
Most RFID tags emit unique identifiers when they respond to reader
interrogation, even tags that protect data with cryptographic
algorithm \cite{Juels2006}. A similar problem arises
for Bluetooth-enabled wireless devices \cite{Jakobsson2001},
as well as for the NFC protocol that is based on the transfer of
an ID for anti-collision during the process of contactless
reading of transponders \cite{Madlmayr2008}.
Although cryptographic techniques can be used to transform
ID data to preserve privacy, thereby complying with the GDPR,
they are computationally-intensive
and require the generation and maintenance of multiple keys,
which also leads to higher energy consumption.
Alternatively, devices' IDs can be aggregated
or perturbed in such a way that individual data privacy
is preserved but, at the same time, useful information
for crowding management is not destroyed.
In the case of WiFi networks, a viable solution
is represented by MAC address randomization \cite{Martin2017},
which is an available option
in different operating systems, such as iOS, Android, and Windows.
However, although it is compliant with the GDPR,
some crowd counting methods are
adversely affected by MAC address randomization.

A cutting-edge future
research direction might be exploiting
the mathematical concept of
differential privacy \cite{Dwork2014}, which
addresses the challenging goal of learning
nothing about an individual while learning useful
information about a population. This
is in accordance with existing data protection laws,
including the GDPR. However,
the feasibility of applying differential privacy approaches
to crowd management remains an open topic of research.
Device-free NVB solutions, both based
on RF or other physical properties (such as audio or ultrasound signals),
are less harmful to an individual's privacy since
they do not require processing of devices' IDs
and, thus, according to the GDPR, they do not collect
personally identifiable information that could potentially
be used to identify a particular person during people counting activities.
This pushes towards emerging NVB techniques using
reflected-power approaches \cite{Yildirim2021}, also referred
to as passive NVB solutions, which reuse existing over-the-air
signals to count people or estimate their flow,
by viewing the signals as power carriers rather
than information ones.


\smallskip
\subsubsection{Communications}

A constant stream of up-to-the-minute data
can help PT system operators stay one step ahead of crowd situations.
However, with limited budgets and compressed time frames, implementing
a state-of-the-art CSS can present fundamental challenges.
There are three basic challenges regarding ITSs based upon
acquisition
and processing of large amounts of sensor data:
bandwidth, latency, and heterogeneity of data and infrastructures.

A first critical issue is represented
by the significant bandwidth demand to connect thousands of devices and
support hundreds of real-time video feeds, along with data gathered from ubiquitous sensors. A potential approach to fulfill such a bandwidth demand is offered by the unlicensed band technology, which has attracted significant effort during the last decade. In particular, the new radio unlicensed (NR-U) technology appears to be particularly suited for ITS applications \cite{Bajracharya2021}, since it considers multiple bands and other deployment scenarios, such as dual connectivity and standalone operation in unlicensed bands.

In addition to bandwidth, since some crowd management
decisions are time-sensitive in nature,
latency is another key performance indicator of the CSS.
Cloud computing at the edge of the network
\cite{Arthurs2021}, namely, close to railway/metro stations, bus stops, and ITS sensors,
can provide a solution for satisfying latency and bandwidth constraints,
thus avoiding unacceptable upload delays
as well as energy consumption of IoT sensors.
In those scenarios, when there is no edge server
nearby that can offload the tasks, employing UAVs is a promising
solution by serving as computing-communications
edge server for resource-constrained IoT devices \cite{Liu2021}.
One can imagine a hybrid communication architecture
for future ITSs where edge nodes locally process the time-critical
raw sensed data, while non-time-sensitive
data are transmitted to the cloud.

From a communication viewpoint,
an ITS is a  heterogeneous network, where a huge number
of devices are connected to global networks
using multiple technologies and platforms
(i.e., cloud, edge, wireless, etc.), with various intelligence levels.
Traditionally, communication between different wireless technologies
is achieved indirectly via gateways equipped with
multiple radio interfaces, which will become a bottleneck
when many heterogeneous IoT devices are deployed.
Cross-technology communications (CTC) opens
a new direction of direct communication among
different wireless technologies,
when they operate in the same spectrum band \cite{Gao2022}.
These technologies are purely a software-based solution,
requiring no hardware modification.

\smallskip
\subsubsection{Crowd control}

A significant research effort is
studying how to optimally employ
(possible heterogenous)
crowding data to optimize (at the various levels)
the PT system behavior.
Suitable performance
metrics must be defined, as well as
optimization methodologies,
especially when applied to large-scale
complex PT networks involving thousands of vehicles
and stations/stops.
Indeed, in many papers crowd control strategies are localized, i.e., they are
applied only to single lines or to a small portion of the network, whereas
obviously crowd control
must be applied holistically to the whole
PT network. As an example, stop-skipping can solve the problem
of a single vehicle overcrowding, but can negatively
affect crowding at the stops or metro platforms.
To perform complex system-wide
optimization,
it seems difficult to apply
model-based approaches. Thus,
the preferred choice should be resorting to techniques such as
data mining, AI, or ML/DL, along
the vision of \textit{data-driven} PT systems
\cite{Zhang2011}.
An interesting issue is to assess the impact
of inaccurate crowd monitoring/prediction
measurements due to sensing
on optimization strategies, in order to devise
robust techniques.
Another interesting development is how to
incorporate crowding data
in existing simulation tools for
transportation planning, like BusMezzo
\cite{Toledo2010} or SUMO \cite{Lopez2018}.
%


\section{Conclusions and directions for future work}
\label{sec:concl}

A clean, smart, and resilient PT system is at
the core of worldwide economies
and is central to people's lives:
this is why there has been an increased
number of research articles that discuss
the feasibility of integrating ICT/IoT
sensing solutions in ITSs.
In this paper, we have reviewed the main
ICT/IoT sensing
technologies for crowd analysis,
showing also how they can be adopted
in a reference architecture
aimed at introducing innovative
crowd management
functionalities
in legacy ITS systems.
The new framework is based
on some basic components and
subsystems, which can be
used as building blocks
to implement an evolved ITS,
capable of real-time
monitoring and predicting crowd situations, as well as
disseminating useful information to users
at the bus stops/stations and/or through mobile
transport apps.
A series of challenges arise from
the proposed reference architecture, such as
incentive mechanisms for user cooperation,
data fusion and processing of heterogeneous data,
security, privacy, broadband and ultra-low latency
communications,  system-wide optimization of crowd control
strategies.

Some features of the new framework
are similar to those of a
contact tracing system, which can be
implemented more easily by resorting
to user cooperation.
In this sense, the crowd detection functionalities
can be incorporated in a more complex system,
which can implement, besides the
typical mobile transport app
functionalities (like Moovit),
also the possibility to buy
tickets and/or to reserve the access to the vehicles,
in conditions of particular crowding.
This could represent a decisive
incentive to the use of a PT system.
However, the more appropriate
crowd monitoring solution must be singled
out case-by-case, in dependence
on the scenario, the ICT infrastructure owned or leased by
the PT operator, the
socio-economic context,
and the cost-benefits ratio.
The potentials of crowd management
go beyond the scope of
dealing with typical social distancing problems,
by also allowing real-time
optimization and adaptive
management of PT systems.

Finally, the study of the open literature makes it clear that an in-depth performance
analysis  of  the sensing techniques for PT systems reported in Table~\ref{tab:SAL-technologies}
has not yet been carried out in each considered scenario.
Therefore, a first interesting research development consists of
performing a comparative statistical
performance evaluation (numerical and/or experimental)
of such sensing techniques
for each outlined indoor/outdoor scenarios.
An additional research issue is to carry
out a performance analysis
of the DF algorithms for
associating, correlating, and combining information
from different sensors, whose accuracy is crucial
to provide effective actions.

\section*{Acknowledgment}

The authors would like to thank the
IEEE ComSoc/VTS Italian Chapter for supporting
this work, by awarding the first prize to the SALUTARY
(Safe and Reliable Public Transportation System)
system concept within the call for ideas in
response to COVID-19 outbreak in Italy.


\printbibliography

\begin{IEEEbiography}
[{\includegraphics[width=1in,height=1.25in,clip,keepaspectratio]{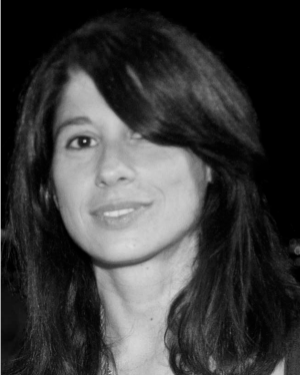}}]
{Donatella Darsena} (M'06-SM'16) received the Dr. Eng. degree summa cum laude in telecommunications engineering in 2001, and the Ph.D. degree in electronic and telecommunications engineering in 2005, both from the University of Napoli Federico II, Italy. From 2001 to 2002, she worked as embedded system designer in the Telecommunications, Peripherals and Automotive Group, STMicroelectronics, Milano, Italy.
Since March 2005, she has been with the University of Napoli Parthenope, Italy.
She first served as an Assistant Professor of probability theory
and, since January 2022, she has served as an Associate Professor of telecommunications with
the Department of Engineering.

Her research interests are in the broad area of signal processing for communications, with current emphasis on backscattering communications, space-time techniques for cooperative and cognitive networks, green communications for IoT.
Dr. Darsena was an Associate Editor for the IEEE COMMUNICATIONS LETTERS from
December 2016 to July 2019.
She has served as Associate Editor for IEEE ACCESS since October 2018,
Area Editor for IEEE COMMUNICATIONS LETTERS
since August 2019, and Associate Editor for
IEEE SIGNAL PROCESSING LETTERS since 2020.
\end{IEEEbiography}

\begin{IEEEbiography}
[{\includegraphics[width=1in,height=1.25in,clip,keepaspectratio]{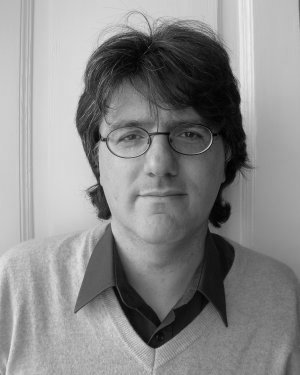}}]
{Giacinto Gelli}(M'18-SM'20)
was born in Napoli, Italy, on July 29, 1964.
He received the Dr. Eng. degree \textit{summa cum laude} in electronic
engineering in 1990, and the Ph.D. degree in computer science and
electronic engineering in 1994, both from the University of Napoli
Federico II.

From 1994 to 1998, he was an Assistant Professor with the
Department of Information Engineering, Second University of
Napoli.
Since 1998 he has been with the Department of Electrical Engineering and Information Technology, University of Napoli Federico II,
first as an Associate Professor,
and since November 2006 as a Full Professor of telecommunications.
He also held teaching positions at the University of Napoli Parthenope.
His research interests are in the broad area of
signal and array processing for communications,
with current emphasis on reflected-power communication systems, 
multicarrier modulation systems, and
space-time techniques for cooperative and cognitive
communications systems.
\end{IEEEbiography}

\begin{IEEEbiography}
[{\includegraphics[width=1in,height=1.25in,clip,keepaspectratio]{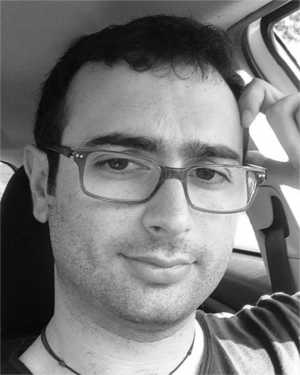}}]
{Ivan Iudice}
was born in Livorno, Italy, on November 23, 1986.
He received the B.S. and M.S. degrees
in telecommunications engineering in 2008 and 2010,
respectively, and the Ph.D. degree
in information technology and electrical engineering in 2017,
all from University of Napoli Federico II, Italy.

Since 2011, he has been with the Italian Aerospace Research Centre (CIRA), Capua, Italy.
He first served as part of the Electronics and Communications laboratory and, since November 2020,
he has served as part of the Security of Systems and Infrastructures laboratory.
His research activities lie in the area of
signal and array processing for communications,
with current interests focused
on physical-layer cyber security
and space-time techniques for
cooperative communications systems.
\end{IEEEbiography}

\begin{IEEEbiography}[
{\includegraphics[width=1in,height=1.25in,clip,keepaspectratio]{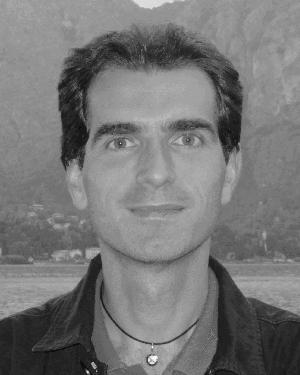}}]
{Francesco Verde}(M'10-SM'14) was born in Santa Maria Capua Vetere,
Italy, on June 12, 1974. He received the Dr. Eng. degree
\textit{summa cum laude} in electronic engineering
from the Second University of Napoli, Italy, in 1998, and the Ph.D.
degree in information engineering
from the University of Napoli Federico II, in 2002.
Since December 2002, he has been with the University of Napoli Federico II, Italy. He first served as an Assistant Professor of signal theory and mobile communications
and, since December 2011, he has served as an Associate Professor of telecommunications with the Department of Electrical Engineering and Information Technology.
His research activities include reflected-power communications,
orthogonal/non-orthogonal multiple-access techniques, wireless systems optimization, and
physical-layer security.

Prof. Verde has been involved in several technical program committees of major IEEE conferences in signal processing and wireless communications.
He has served as Associate Editor for IEEE TRAN\-SACTIONS ON VEHICULAR TECHNOLOGY since 2022 and Senior Area Editor of the IEEE SIGNAL PROCESSING LETTERS since 2018.
He was an Associate Editor of the IEEE TRANSACTIONS ON SIGNAL PROCESSING (from 2010 to 2014), IEEE SIGNAL PROCESSING LETTERS (from 2014 to 2018), and
IEEE TRANSACTIONS ON COMMUNICATIONS (from 2017 to 2022),
as well as Guest Editor of the EURASIP Journal on Advances in Signal Processing in 2010 and SENSORS MDPI in 2018-2022.
\end{IEEEbiography}

\end{document}